\documentclass[aps,showpacs]{revtex4}

\usepackage{graphicx}
\usepackage{amsmath}
\usepackage{amssymb}

\begin{document}

%\preprint{HEP/123-qed}

\title[Finite Temperature Time-Dependent]{Finite Temperature 
Time-Dependent Effective Theory  for \\ 
the Phase Field in two-dimensional $d$-wave Neutral Superconductor}

\author{S.G.~Sharapov} \thanks{On leave of absence from Bogolyubov
 Institute for Theoretical Physics, Kiev, Ukraine}%
 \email{Sergei.Sharapov@unine.ch} \author{H.~Beck}%
 \email{Hans.Beck@unine.ch} 
 \affiliation{Institut de Physique,
 Universit\'e de Neuch\^atel, 2000 Neuch\^atel, Switzerland}
 \author{V.M.~Loktev} \email{vloktev@bitp.kiev.ua}
 \affiliation{ Bogolyubov Institute for Theoretical Physics,
 Metrologicheskaya Str. 14-b, Kiev, 03143, Ukraine}

\date{March 15, 2001}% It is always \today, today, but you may specify any
% date with \date.

\begin{abstract}
We derive  finite temperature time-dependent effective actions
for the phase of the pairing field, which are appropriate for a 
2D electron system with both non-retarded $d$- and $s$-wave attraction. 
As for $s$-wave pairing the $d$-wave effective action contains terms 
with Landau damping,
but their structure appears to be different from the s-wave case 
due to the fact that the Landau damping is determined by the
quasiparticle group velocity $\mathbf{v}_{g}$, which for $d$-wave
pairing does not have the same direction as the non-interacting 
Fermi velocity $\mathbf{v}_{F}$. We show that for $d$-wave pairing the Landau
term has a linear low temperature dependence and in contrast
to the $s$-wave case are important for all finite temperatures. 
A possible experimental observation of the phase excitations is discussed.
\end{abstract}


\maketitle

\section{Introduction}
The microscopic derivation of the effective time-dependent Ginzburg-Landau
(GL) theory continues to attract attention since an early paper
by Abrahams and Tsuneto \cite{Abrahams}. Whereas the static GL potential
was derived \cite{Gor'kov} from the microscopic BCS theory soon after its 
introduction, the time-dependent GL theory is still a subject of interest
(see \cite{Aitchison.1997,Aitchison} for a review on the problem's history).
One of the reasons for this is the presence of Landau damping terms
in the effective action. For $s$-wave superconductivity these terms
are singular at the origin of energy-momentum space, and consequently
they cannot be expanded as a Taylor series about the origin.
In other words, these terms do not have a well-defined expansion
in terms of space and time derivatives of the ordering field and therefore
they cannot be represented as a part of a local Lagrangian.
We recall that at $T =0$ and for the static (time-independent) case
the Landau damping vanishes, so that either at $T =0$ one still has
a local well-defined time-dependent GL theory or for $T \neq 0$
the familiar static GL theory exists. It is known, however,
that for $s$-wave superconductivity even though the Landau terms
do exist, they appear to be small compared to the main terms of
the effective action in the large temperature region 
$0 < T \lesssim 0.6 T_c$
\cite{Aitchison}, where $T_c$ is the superconducting transition temperature. 
This is evidently related to the fact that only
thermally excited quasiparticles  contribute to the Landau damping.
The number of such quasiparticles at low temperatures appears to be a small
fraction of the total charge carriers number in the $s$-wave
superconductor due to the nonzero superconducting gap
$\Delta_s$ which opens over all directions on the Fermi surface.

For a $d$-wave superconductor there are four Dirac points (nodes)
where the superconducting gap $\Delta_{d}({\bf k})$ becomes zero
on the Fermi surface. 
The presence of the nodes increases significantly the number of the 
thermally excited quasiparticles at given temperature $T$ comparing to 
the $s$-wave case. Therefore one can expect that the Landau damping 
should be stronger for superconductors with a $d$-wave gap which 
is commonly accepted to be the case of  high-temperature 
superconductors (HTSC) \cite{d-wave}. Moreover, it is believed that
at temperatures $T \ll T_c$,these quasiparticles are reasonably
well described by the Landau quasiparticles, even though such an
approach fails in these materials at higher energies \cite{Lee}. 
This is the reason why one can hope that a generalization of the BCS-like
approach \cite{Aitchison} for the 2D $d$-wave superconductivity may be
relevant to the description of the low-temperature time-dependent
GL theory in HTSC.

In this work we derive such a theory from a microscopical model with
$d$-wave pairing extending the approach of \cite{Aitchison} 
developed for $s$-wave superconductivity. As known from \cite{Abrahams}
the physical origin of the Landau damping is a scattering of the
thermally excited quasiparticles (``normal'' fluid) with  group
velocity $\mathbf{v}_g$ from the excitations of  phase 
(or $\theta$-) quantums. Such 
conversion occurs only if the {\v C}erenkov condition,
$\Omega = \mathbf{v}_g \mathbf{K}$ for the energy $\Omega$ and momentum
$\mathbf{K}$ of the $\theta$-excitation is satisfied. This phenomenon
in superconductivity is also called Landau damping since its
equivalent in the plasma theory  was originally obtained by Landau 
(see e.g. \cite{Sitenko}).

To emphasize the difference between the Landau damping for $s$- and
$d$-wave pairing, we also derive for the comparison the corresponding terms 
for a 2D $s$-wave superconductor. In addition we compare 2D expressions
obtained here with the 3D $s$-wave case studied in \cite{Aitchison}.
The collective phase oscillations in charged $d$-wave superconductor
for clean and dirty cases were recently studied in \cite{Takada}.
Due to the complexity of the corresponding equations they were solved 
numerically, neglecting damping of the phase excitations. Thus our fully
analytical treatment can be very useful for further studies
of the phase excitations. We also mention a recent paper 
\cite{Randeria.action} where the effective action for the phase mode in 
the $d$-wave superconductor was obtained using the cumulant expansion. 
The Landau terms were neglected in \cite{Randeria.action}, but the effect 
of Coulomb interaction was taken into account.   

Our main results can be summarized as follows.

1. We find that the main physical difference between the $s$-
and $d$-wave cases is related to the fact that for $d$-wave 
superconductivity the direction of the quasiparticle group
velocity $\mathbf{v}_{g}(\mathbf{k})\equiv \partial E(\mathbf{k})/\partial \mathbf{k}$ ($E(\mathbf{k})$ is the quasiparticle dispersion law) does not
coincide with the Fermi velocity $\mathbf{v}_F$ 
\cite{Lee,Carbotte} and a gap velocity  
$\mathbf{v}_{\Delta} \equiv \partial \Delta_{d}(\mathbf{k})/\partial \mathbf{k}$
also enters into the {\v C}erenkov condition along with $\mathbf{v}_F$.

2. We show that the intensity of the Landau damping has a linear
temperature dependence at low $T$ with a coefficient expressed
in terms of the anisotropy $\alpha_{D} \equiv v_F/ v_{\Delta}$ of
the Dirac spectrum $E(\mathbf{k}) = \sqrt{v_F^2 k_1^2 + v_{\Delta}^2 k_2^2}$.
Here $k_1$ ($k_2$) are the projections of the quasiparticle momentum
on the directions perpendicular (parallel) to the Fermi surface.
The parameters $v_{F}$, $v_{\Delta}$ and $\alpha_{D}$ proved to be very
convenient both in the theory, for example,  of the transport phenomena 
\cite{Lee}, ultrasonic attenuation \cite{Carbotte} in $d$-wave superconductors
and for the analysis of various experiments \cite{Chiao}. 

3. We find that the Landau damping is sensitive to the direction
of the phason momentum. In particular, for a given node the Landau
damping is possible only if the components of the phason momentum 
${\bf K} = (K_1, K_2)$ (which are defined exactly as the components
of the quasiparticle momentum above) satisfy the
condition $|\Omega| < \sqrt{v_{F}^{2} K_{1}^{2} + v_{\Delta}^{2} K_{2}^{2}}$.

4. We derive a simple approximate representation for the Landau
damping terms which can be useful for further studies of the
$d$-wave superconductors.  

5. We derive an approximate expression for the propagator of the
Bogolyubov-Anderson mode which includes the Landau damping.

6. Concerning the mathematical formalism used in the paper,
we adapt the bilocal Hubbard-Stratonovich field method of \cite{Kleinert}
to the $d$-wave pairing. Additionally we adjust the technique of the 
derivative expansion for the ``phase only'' action 
(see \cite{Aitchison} and Refs. therein) for the model with the
tight-binding spectrum.

The paper is organized as follows:
In Section~\ref{sec:model} we present our model and write down
the partition function using a bilocal Hubbard-Stratonovich field.
In Section~\ref{sec:action}, we introduce the modulus-phase variables and
represent the effective ``phase only'' action as an infinite series.
It appears that the low-energy phase dynamics is contained in the first two 
terms which are evaluated, respectively,
in Sections~\ref{sec:first.tr} and \ref{sec:second.tr} with some
details considered in Appendix~\ref{sec:A}.
The effective Lagrangians for the $d$- and $s$-wave cases without the Landau
damping are discussed in Section~\ref{sec:local.action}.
In Section \ref{sec:Landau.damping} we derive the damping terms
and in detail compare $d$- and $s$-wave cases. The approximate forms
of the effective action and $\theta$- propagator are considered in 
Section~\ref{sec:approximate}. Section~\ref{sec:conclusion}    
presents our conclusions and comments on a possible experimental
observation of the phase excitations.

\section{Model}
\label{sec:model}
Let us consider the following action
\begin{equation}
S = - \int_{0}^{\beta} d \tau \left[\sum_{\sigma} \int
d^2r \psi_{\sigma}^{\dagger}(\tau, {\bf r})
\partial_{\tau} \psi_{\sigma}(\tau, {\bf r}) + H(\tau) \right]\,,
\qquad \mathbf{r} = (x,y)\,, 
\qquad \beta \equiv \frac{1}{T}\,,\label{S}
\end{equation}
where the Hamiltonian $H(\tau)$ is
\begin{equation}
\label{Hamilton}
\begin{split}
H(\tau) & = \sum_{\sigma} \int d^2 r \psi_{\sigma}^{\dagger}(\tau, {\bf r})
[\varepsilon(- i \nabla) - \mu] \psi_{\sigma}(\tau, {\bf r})      \\
& - \frac{1}{2} \sum_{\sigma} \int d^2 r_1 \int d^2 r_2
\psi_{\sigma}^{\dagger}(\tau, {\bf r}_2) \psi_{\bar{\sigma}}^{\dagger}(\tau,
{\bf r}_1) V({\bf r}_1; {\bf r}_2) \psi_{\bar{\sigma}}(\tau, {\bf r}_1)
\psi_{\sigma}(\tau, {\bf r}_2)\,.
\end{split}
\end{equation}
Here  $\psi_{\sigma}(\tau, {\bf r})$ is a fermion field with the spin 
$\sigma= \uparrow, \downarrow$, $\bar{\sigma} \equiv - \sigma$,
$\tau$ is the imaginary time and $V({\bf r}_1; {\bf r}_2)$
is an attractive potential.
For the sake of the simplicity we consider the dispersion law,
$\varepsilon(\mathbf{k}) = -2t (\cos k_x a + \cos k_y a )$ for a model on a
square lattice with the constant $a$ including the nearest-neighbor
hopping $t$ only. This, however, is not an essential restriction because the 
final results for the $d$-wave case will be formulated in terms of 
the non-interacting Fermi velocity 
$\mathbf{v}_F \equiv \partial \varepsilon(\mathbf{k})/\partial 
\mathbf{k}|_{\mathbf{k} = \mathbf{k}_F}$ 
and the gap velocity 
$\mathbf{v}_\Delta$ defined in the Introduction.

The bilocal Hubbard-Stratonovich fields 
$\Phi (\tau, \mathbf{r}_1; \mathbf{r}_2)$
and  $\Phi^{\dag}(\tau, \mathbf{r}_1; \mathbf{r}_2)$
(see e.g. \cite{Kleinert}) can be
utilized to study the model (\ref{S}), (\ref{Hamilton})
\begin{equation}
\label{Hubbard}
\begin{split}
& \exp\left[\int_{0}^{\beta} d \tau \int d^2 r_1 \int d^2 r_2
\psi_{\uparrow}^{\dagger}(\tau, {\bf r}_2) \psi_{\downarrow}^{\dagger}(\tau,
{\bf r}_1) V(\mathbf{r}_1; \mathbf{r}_2) \psi_{\downarrow}(\tau, {\bf r}_1)
\psi_{\uparrow}(\tau, {\bf r}_2)\right] \\
& = \int \mathcal{D} \Phi^{\dag}(\tau, \mathbf{r}_1; \mathbf{r}_2) \mathcal{D}
\Phi (\tau, \mathbf{r}_1; \mathbf{r}_2)
\exp \left[- \int_{0}^{\beta} d \tau \int d^2 r_1 \int d^2 r_2
\frac{1}{V(\mathbf{r}_1; \mathbf{r}_2)}
|\Phi(\tau,\mathbf{r}_1; \mathbf{r}_2)|^2\right.\\
&\left. +\int_{0}^{\beta} d \tau \int d^2 r_1 \int d^2 r_2
(\Phi^{\dag}(\tau,\mathbf{r}_1; \mathbf{r}_2)\psi_{\downarrow}(\tau, {\bf r}_1)
\psi_{\uparrow}(\tau, {\bf r}_2)+ 
\psi_{\uparrow}^{\dagger}(\tau, {\bf r}_1)
\psi_{\downarrow}^{\dagger}(\tau, {\bf r}_2)
\Phi (\tau, \mathbf{r}_1; \mathbf{r}_2))
\right]\,.
\end{split}
\end{equation}
On the right hand side, $1/V(\mathbf{r}_1; \mathbf{r}_2)$ is understood as
numeric division, no matrix inversion being implied. The hermitian conjugate of
$\Phi(\tau,\mathbf{r}_1; \mathbf{r}_2)$ includes the transpose in the
functional sense, i.e. $\Phi^{\dag}(\tau,\mathbf{r}_1; \mathbf{r}_2) \equiv
[\Phi(\tau,\mathbf{r}_2; \mathbf{r}_1)]^{\ast}$.

Thus in the Nambu variables
\begin{equation}
\label{Nambu.variables}
\Psi(\tau, \mathbf{r}) = \left( \begin{array}{c} 
\psi_{\uparrow}(\tau, \mathbf{r}) \\
\psi_{\downarrow}^{\dagger}(\tau, \mathbf{r})
\end{array} \right), \qquad
\Psi^{\dagger}(\tau, \mathbf{r}) = \left( \begin{array}{cc} 
\psi_{\uparrow}^{\dagger}(\tau, \mathbf{r})
\quad \psi_{\downarrow}(\tau, \mathbf{r})
\end{array} \right)
\end{equation}
the partition function can be written as
\begin{equation}
\label{statistical.sum}
\begin{split}
Z = & \int \mathcal{D} \Psi^{\dagger} \mathcal{D} \Psi \mathcal{D} \Phi^{\dag}
\mathcal{D} \Phi \exp \left\{\int_{0}^{\beta} d \tau \int d^2 r_1 \int d^2 r_2
\left[-\frac{1}{V(\mathbf{r}_1; \mathbf{r}_2)}
|\Phi (\tau,\mathbf{r}_1; \mathbf{r}_2)|^2 \right. \right. \\
& + \Psi^{\dagger}(\tau,\mathbf{r}_1)
(-\partial_{\tau}-\tau_{3}\xi(-i\tau_{3}\nabla))\Psi(\tau,\mathbf{r}_2)
\delta(\mathbf{r}_1 - \mathbf{r}_2) \\
& \left. \left. + \Phi^{\dag}(\tau,\mathbf{r}_1; \mathbf{r}_2)
\Psi^{\dagger}(\tau,\mathbf{r}_1) \tau_{-}\Psi(\tau,\mathbf{r}_2) +
\Psi^{\dagger}(\tau,\mathbf{r}_1) \tau_{+}\Psi(\tau,\mathbf{r}_2)
\Phi(\tau,\mathbf{r}_1; \mathbf{r}_2) \right] \right\}
\end{split}
\end{equation}
where $\xi(-i \tau_3 \nabla) \equiv \varepsilon(-i \tau_3 \nabla) - \mu$
and $\tau_3$, $\tau_{\pm} = (\tau_1 \pm i \tau_{2})/2$ are Pauli matrices.

In general an electron-electron attraction on the nearest-neighbor
lattice sites can be considered (see, for example, \cite{Nikolaev,Meintrup}).
The momentum representation for this interaction  contain in the 
pairing channel extended $s$-, $d$- and even $p$-wave pairing terms:
\begin{equation}
\label{all.pairings}
\begin{split}
& V(\mathbf{k} - \mathbf{k}^\prime)  = V
[\cos(k_x - k_x^{\prime}) a + \cos(k_y - k_y^{\prime}) a] \\
& = \frac{V}{2}  
(\cos k_x a + \cos k_y a)(\cos k_x^\prime a + \cos k_y^\prime a) 
 + \frac{V}{2}
(\cos k_x a - \cos k_y a)(\cos k_x^\prime a - \cos k_y^\prime a) \\
& + V (\sin k_x \sin k_x^{\prime} + \sin k_y \sin k_y^{\prime}).
\end{split}
\end{equation}
Motivating by
HTSC we consider here $d$-wave pairing only, so 
that for the Fourier transform of the pairing potential
$V(\mathbf{r}_1 - \mathbf{r}_2)$ we use 
\begin{equation}
\label{Pairing.potential} 
V(\mathbf{k} - \mathbf{k}^\prime) = V_d (\cos k_x a -
\cos k_y a)(\cos k_x^\prime a - \cos k_y^\prime a)\,.
\end{equation}

As was mentioned in Introduction, we will compare the main results
for the $d$-wave case with the simplest 2D continuum $s$-wave
pairing model (see e.g. \cite{Gusynin.JETP1} and the
review \cite{Loktev.review}) which has a quadratic dispersion
law $\varepsilon({\bf k}) = {\bf k}^2/2m$ and a local attraction
$V({\bf r}_1 - {\bf r}_2) = V \delta ({\bf r}_1 - {\bf r}_2)$.

\section{The Effective Action}
\label{sec:action}

While for the model with the local four-fermion attraction 
\cite{Aitchison,Gusynin.JETP1}
the modulus-phase
variables could be introduced exactly, one should apply an additional
approximation for treating the present model.

Let us split the charged fermi-fields $\psi(\tau, \mathbf{r})$ and
$\psi^{\dagger}(\tau, \mathbf{r})$ in (\ref{statistical.sum}) into the neutral
fermi-field $\chi(\tau, \mathbf{r})$ \cite{foot1}
and charged Bose-field $\exp[i \theta(\tau, \mathbf{r})/2]$
\begin{equation}\label{fermi.parametrization}
\psi_{\sigma}(\tau, \mathbf{r}) = \chi_{\sigma}(\tau, \mathbf{r}) \exp [i
\theta(\tau, \mathbf{r})/2]\,, \qquad \psi_{\sigma}^{\dagger}(\tau,
\mathbf{r}) = \chi_{\sigma}^{\dagger}(\tau, \mathbf{r}) \exp [-i \theta(\tau,
\mathbf{r})/2]\,.
\end{equation}
It is clear that the terms containing $\delta(\mathbf{r}_1 - \mathbf{r}_2)$ in
(\ref{statistical.sum}) can be treated  similarly to the old model and the
problem arises when one deals with the Hubbard-Stratonovich field. To consider
this field we introduce the relative $\mathbf{r} = \mathbf{r}_1 - \mathbf{r}_2$
and center of mass coordinates $\mathbf{R} = (\mathbf{r}_1 + \mathbf{r}_2)/2$.
%This, of course, implies that $\mathbf{r}_1 = \mathbf{R} + \mathbf{r}/2$ and
%$\mathbf{r}_2 = \mathbf{R} - \mathbf{r}/2$ for a pair of identical particles.
Now we can introduce the modulus-phase representation for the
Hubbard-Stratonovich field
\begin{equation}\label{Phi.modulus}
\Phi(\tau, \mathbf{r}_1, \mathbf{r}_2) \equiv 
\Phi(\tau, \mathbf{R}, \mathbf{r}) =
\Delta(\tau, \mathbf{R}, \mathbf{r}) 
\exp[i \theta(\tau, \mathbf{R}, \mathbf{r})]\,,
\end{equation}
where $\Delta(\mathbf{R}, \mathbf{r})$ is the modulus of the Hubbard-Stratonovich
field  and $\theta(\mathbf{R}, \mathbf{r})$ is its phase.

Assuming that the global phase $\theta(\mathbf{R}, \mathbf{r})$ 
varies slowly over distances on the order of a Cooper pair size and thus
is not sensitive to the inner pair structure described by the relative
variable $\mathbf{r}$, we can rewrite (\ref{Phi.modulus}) as
\begin{equation}\label{Phi.modulus.approx}
\Phi(\tau, \mathbf{R}, \mathbf{r}) \approx \Delta(\tau, \mathbf{R}, \mathbf{r}) 
\exp[i \theta(\mathbf{\tau, R})]\,.
\end{equation}
The approximation we made writing Eq.~(\ref{Phi.modulus.approx}) is in fact
equivalent to the Born-Oppenheimer approximation \cite{Thouless.1993}
which allows one to separate the dynamics of the Cooper pair formation 
described by the relative coordinate ${\bf r}$ in 
$\Delta(\tau, \mathbf{R},\mathbf{r})$
from the motion of the superconducting condensate
described by the center mass coordinate ${\bf R}$ in 
$\theta(\tau, \mathbf{R})$. 
If the condensate motion 
is slow enough this separation becomes
possible because the dynamics of the Cooper pair formation
can always follow the motion of the condensate. 
Using the lattice language one can also say about
(\ref{Phi.modulus.approx}) that the {\it bond} phase
is replaced by the {\it site} phase \cite{Tesanovic}.

Applying the transformations (\ref{Phi.modulus.approx}) and
(\ref{fermi.parametrization}) to the terms with the Hubbard-Stratonovich field
in (\ref{Hubbard}) we obtain (the imaginary time $\tau$ is omitted)
\begin{equation}\label{interact}
\begin{split} 
& \Phi^{\dag}(\mathbf{R}; \mathbf{r}) \psi_{\downarrow}( {\bf r}_1)
\psi_{\uparrow}({\bf r}_2) = 
\Phi^{\dag}(\mathbf{R}; \mathbf{r}) 
\chi_{\downarrow}( {\bf r}_1) 
\exp \left[\frac{i \theta (\mathbf{R} + \mathbf{r}/2)}{2}\right]
\chi_{\uparrow}({\bf r}_2)  
\exp \left[\frac{i \theta (\mathbf{R} - \mathbf{r}/2)}{2}\right]
\\ 
& \approx \Delta(\mathbf{R}; \mathbf{r}) \exp[- i
\theta(\mathbf{R})] \chi_{\downarrow}( {\bf r}_1) \chi_{\uparrow}({\bf r}_2)
\exp \left[i \theta(\mathbf{R}) + \frac{r_\alpha}{2} \frac{r_{\beta}}{2}
\nabla_{\alpha} \nabla_{\beta} \theta(\mathbf{R})\right]
\approx \Delta(\mathbf{R}; \mathbf{r}) \chi_{\downarrow}( {\bf r}_1)
\chi_{\uparrow}({\bf r}_2)\,,
\end{split}
\end{equation}
where we used the assumption (or
hydrodynamical, long wavelength approximation, see
\cite{Palo.1999}) that $\theta(\mathbf{R})$ varies slowly,
%\begin{equation}\label{hydrodynamic}
$\theta(\mathbf{R}) \gg 
%r_{\alpha} r _{\beta} \nabla_{\alpha} \theta(\mathbf{R}) 
%\nabla_{\beta} \theta(\mathbf{R}) \approx 
\xi_0^2 (\nabla \theta(\mathbf{R}))^2$.
%\end{equation}
Here $\xi_0$ is the coherence length which for the BCS theory coincides
with an average pair size.

Then the partition function in the modulus-phase variables is
\begin{equation}
\label{statistical.sum.phase} Z = \int \Delta \mathcal{D} \Delta 
\mathcal{D} \theta
\exp[-\beta \Omega(\Delta, \partial \theta) ]\,,
\end{equation}
where the effective potential
\begin{equation}
\beta \Omega(\Delta, \partial \theta) = \int_{0}^{\beta} d \tau \int d^2 r_1
\int d^2 r_2 \frac{\Delta^{2}(\tau,\mathbf{R},\mathbf{r})}{V(\mathbf{r}_1 -
\mathbf{r}_2)} - \mbox{TrLn} G^{-1}  \label{Omega}
\end{equation}
with
\begin{equation}
\label{Green}
G^{-1} = \mathcal{G}^{-1} - \Sigma\,,
\end{equation}
\begin{equation}
\label{Green.neutral}
\begin{split}
& \mathcal{G}^{-1}(\tau_1,\tau_2; \mathbf{r}_1, \mathbf{r}_2) \equiv \langle
\tau_1, \mathbf{r}_1|\mathcal{G}^{-1}| \tau_2, \mathbf{r}_2 \rangle \\ & =
\left[ -\hat{I} \partial_{\tau_{1}} - \tau_{3} \xi(-i \tau_{3}
\nabla_{1})\right] \delta(\tau_1 - \tau_2) \delta(\mathbf{r}_1 - \mathbf{r}_2)
+ \tau_1 \Delta(\tau_1 - \tau_2, \mathbf{R},\mathbf{r})\,;
\end{split}
\end{equation}
\begin{equation}
\label{Sigma}
\begin{split}
& \langle \tau_1, \mathbf{r}_1|\Sigma| \tau_2, \mathbf{r}_2 \rangle = \left[
\tau_{3} \left(i \frac{\partial_{\tau_{1}} \theta}{2} + t a^2
\frac{(\nabla_{x_{1}} \theta)^2}{4} \cos(-i a \nabla_{x_{1}}) 
+ (x \to y) \right) \right.
\\& \left. + \hat{I}\left( - \frac{i t a^2 \nabla_{x_{1}}^2 \theta }{2} 
\cos(-i a \nabla_{x_{1}}) + ta \nabla_{x_{1}} \theta \sin(-i a \nabla_{x_{1}}) 
+ (x \to y) \right)\right] 
\delta(\tau_1 - \tau_2) \delta(\mathbf{r}_1 - \mathbf{r}_2)\,.
\end{split}
\end{equation}
Thus the gauge transformation (\ref{fermi.parametrization})
resulted in the separation of the dependences on $\Delta$ and $\theta$,
viz. $\theta$ is present only in $\Sigma$. The similar method of
the derivative expansion was used before in \cite{Aitchison.1995,Aitchison}.
As pointed out in \cite{Aitchison} the method allows to
maintain explicitly the Galilean invariance 
(the Landau terms break it) and the continuity equation, while the 
expansion $\Phi(x) = \Delta + \Phi_{1}(x) + i \Phi_{2}(x)$ used recently 
in  \cite{Takada} demands the additional 
enforcement of the conservation laws \cite{Kulik}. 

Since the low-energy dynamics in the phase in which
$\Delta \neq 0$ is determined by the long-wavelength fluctuations
of $\theta(x)$, only the lowest order derivatives of the phase 
such as $\nabla \theta$, $\partial_{\tau} \theta$ and $\nabla^2 \theta$ 
need be retained in what follows. However, to take into account
the tight-binding electron spectrum the operators $\sin(-ia \nabla)$
and $\cos(-i a \nabla)$ must be kept. Thus in $\Sigma$ we have omitted 
higher order terms in $\nabla \theta$, but in order to keep all relevant 
terms in the expansion of $\sin (-i a \nabla)$ the necessary resummation 
was done \cite{Sharapov:1998:PC}. One can easily see that for the quadratic
dispersion law $2 t \to 1/(m a^2)$, $\cos (-i a \nabla) \to 1$
and $\sin(-i a \nabla) \to - i a \nabla$, so that Eq.~(\ref{Sigma})
reduces to the known expression from \cite{Aitchison.1995,Aitchison}.
Thus we arrive at the one-loop effective action
\begin{equation}
\Omega \simeq
\Omega _{kin} (v, \mu, T, \Delta, \partial \theta) +
\Omega _{pot}^{\mbox{\tiny MF}} (v, \mu, T, \Delta)
                  \label{kinetic.phase+potential}
\end{equation}
where
\begin{equation}
\Omega _{kin} (\mu, T, \Delta, \partial \theta)
 =  T \mbox{Tr} \sum_{n=1}^{\infty}
\left. \frac{1}{n} ({\cal G} \Sigma)^{n}
\right|_{\partial \Delta/\partial \mathbf{R}=0}
               \label{Omega.Kinetic.phase}
\end{equation}
and
\begin{equation}
\Omega _{pot}^{\mbox{\tiny MF}} (\mu, T, \Delta)  =
\left. \left(  \int d^2  R \int d^2 r  
\frac{\Delta^{2}(\mathbf{r})}{V(\mathbf{r})} -
T \mbox{Tr Ln} {\cal G}^{-1} \right) 
\right|_{\partial \Delta/\partial \mathbf{R}=0}.
                 \label{Omega.Potential.modulus}
\end{equation}
Deriving the ``phase only'' action for the $s$-wave model it was possible to
use $\Delta(\mathbf{R}, \mathbf{r}) = \mbox{const}$ \cite{Loktev.review}. 
The $d$-wave case is more complicated because one should keep the dependence 
on the relative coordinate, 
$\Delta(\mathbf{R}, \mathbf{r}) \approx \Delta(\mathbf{r})$ which is related to
the nontrivial pairing. 
The dependences of the gap $\Delta$ on $T$, $\mu$ and $V_d$ follow
from the extremum condition for the mean-field  
($\partial \Delta /\partial \mathbf{R} =0$) potential
$\partial \Omega_{pot}^{\mbox{\tiny MF}} /\partial \Delta =0$
which results in the usual BCS gap equation.  
For the $d$-wave pairing potential (\ref{Pairing.potential}) 
one obtains
\begin{equation}\label{rho}
 \Delta_d(\mathbf{k})= \frac{\Delta_{d}}{2}(\cos k_x a - \cos k_y a)\,,
\end{equation}
where $\Delta_d$ is the gap amplitude. In our case there is no need 
to solve the gap equation and express $\Delta_d$ in terms of
$T$, $\mu$ and $V_d$ since in what follows we will use $\Delta_d$,
or more precisely the velocity $\mathbf{v}_{\Delta}$, as the input
parameters and will be interested in the low temperature 
($T \ll \Delta_d $) regime.

%It is clear that in contrast to the $s$-wave case where $\Delta \geq 0$, the
%$d$-wave modulus $\Delta(\mathbf{k})$ may also be negative. 
Thus assuming that
$\Delta(\mathbf{R},\mathbf{r})$ does not depend on $\mathbf{R}$ one obtains for
the frequency-momentum representation of (\ref{Green.neutral})
\begin{equation}\label{Green.neutral.momentum}
\mathcal{ G}(i \omega_{n},  {\bf k}) = - \frac{ i \omega_{n} \hat{I} +
\tau_{3} \xi( {\bf k}) - \tau_{1} \Delta(\mathbf{k})} {\omega_{n}^{2} + \xi^{2}(
{\bf k}) + \Delta^{2}(\mathbf{k})}\,,
\end{equation}
where $\Delta(\mathbf{k})$ is given by (\ref{rho}) and
$\omega_n = \pi (2n+1)T$ is fermionic (odd) Matsubara frequency.

The phase dynamics is contained in the kinetic part $\Omega_{kin}$
of the effective action which only involves the single degree of freedom
$\theta$. As discussed in \cite{Aitchison}, it is enough to restrict
ourselves to terms with $n=1,2$ in the infinite series in 
(\ref{Omega.Kinetic.phase}) since at $T=0$ this would give
the right answer for a local time-dependent GL functional  which 
involves the derivatives not higher than $(\nabla \theta)^4$ and
$(\partial_t \theta)^2$.

\section{The First Order Term of the Effective Action and 
the Nodal Approximation}
\label{sec:first.tr}

In this section we calculate the first ($n=1$) term of the sum appearing
in (\ref{Omega.Kinetic.phase}):
\begin{equation}\label{kinetic.1}
\begin{split}
& \Omega_{\mathrm{kin}}^{(1)}  = T \mbox{Tr} [\mathcal{G} \Sigma] \\
& = T \int_{0}^{\beta} d \tau \int d^2 r \left\{T \sum_{n=-\infty}^{\infty}
\int \frac{d^2 k}{(2 \pi)^2} \mbox{tr} [\mathcal{G}(i \omega_n,
\mathbf{k})\tau_3] \left(i \frac{\partial_{\tau} \theta}{2} + \frac{ta^2}{4}
(\nabla_x \theta)^2 \cos k_x a + 
\frac{ta^2}{4} (\nabla_y \theta)^2  \cos k_y a  \right) \right\}\,.
\end{split}
\end{equation}
Summing over Matsubara frequencies, one obtains
\begin{equation}\label{kinetic.1.final}
\Omega_{\mathrm{kin}}^{(1)}  = T \int_{0}^{\beta} d \tau \int d^2 r \left[ \int
\frac{d^2 k}{(2 \pi)^2} n(\mathbf{k})\left( i \frac{\partial_{\tau} \theta}{2}
+ m_{xx}^{-1}(\mathbf{k})\frac{(\nabla_x \theta)^2}{8} +
m_{yy}^{-1}(\mathbf{k}) \frac{(\nabla_y \theta)^2}{8} \right) \right]
\end{equation}
with $m_{xx}^{-1}(\mathbf{k})\equiv \partial^2 \xi(\mathbf{k})/\partial k_x^2$,
$m_{yy}^{-1}(\mathbf{k})\equiv \partial^2 \xi(\mathbf{k})/\partial k_y^2$ and
\begin{equation}\label{n(k)}
n(\mathbf{k}) = 1 - \frac{\xi(\mathbf{k})}{E(\mathbf{k})} \tanh
\frac{E(\mathbf{k})}{2T}\,, \qquad E(\mathbf{k}) =
\sqrt{\xi^2(\mathbf{k})+\Delta^2(\mathbf{k})}\,.
\end{equation}

For $T \ll \Delta_{d}$
linearizing the quasiparticle spectrum about the nodes and defining a
coordinate system $(k_1, k_2)$ at each node with $\hat{\mathbf{k}}_1$
($\hat{\mathbf{k}}_2$) perpendicular (parallel) to the Fermi surface, we can
replace the  momentum integration in (\ref{kinetic.1.final}) by an integral
over the $\mathbf{k}$-space area surrounding each node \cite{Lee}. 
If we further define a
scaled momentum $\mathbf{p} = (p_1, p_2) = (p, \varphi)$ we can let
\begin{equation}\label{nodal.approximation}
\int \frac{d^2 k}{(2 \pi)^2} \to \sum_{j =1}^{4}\int \frac{d k_1 d k_2}{(2
\pi)^2} \to \sum_{j =1}^{4}\int \frac{d^2 p}{(2 \pi)^2 v_{F} v_\Delta} =
\sum_{j =1}^{4}\int_{0}^{p_{max}} \frac{p d p}{2 \pi v_{F} v_\Delta}
\int_{0}^{2 \pi} \frac{d \varphi}{2 \pi}\,, \quad p_{max} = \sqrt{\pi v_{F}
v_{\Delta}}\,,
\end{equation}
where $p_1 \equiv v_{F} k_1 = \varepsilon(\mathbf{k}) = p \cos \varphi$, $p_2
\equiv  v_{\Delta} k_2 = \Delta_d(\mathbf{k}) = p \sin \varphi$ and $p =
\sqrt{p_1^2 + p_2^2} = \sqrt{v_{F}^2 k_1^2 + v_\Delta^2 k_2^2} =
E(\mathbf{k})$. Note that for the particular square lattice model used above
those velocities are $v_{F}= 2 \sqrt{2}ta$ and $v_{\Delta} = \Delta_{d}
a/\sqrt{2}$ respectively.

\begin{figure}[h]
\centering{
\includegraphics[width=13cm]{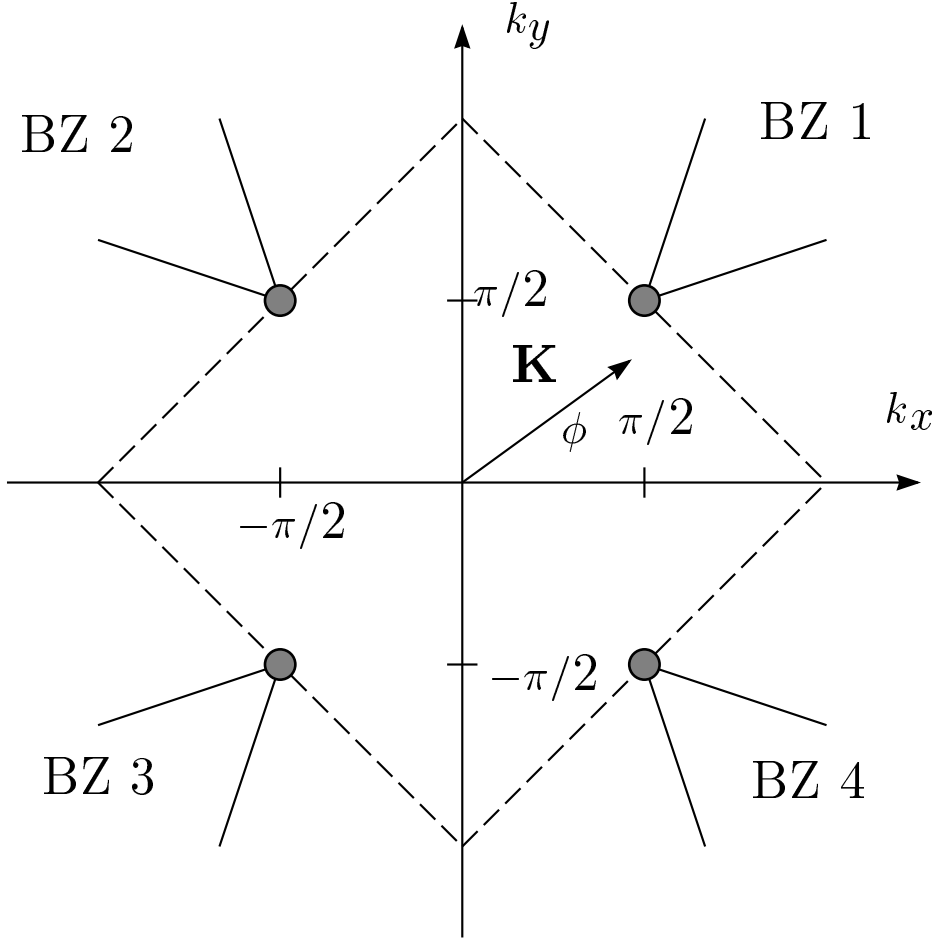}}
\caption{For $\mu =0$ the Fermi surface (dotted line) represents
the points where $\varepsilon (\mathbf{k}) = -2t (\cos k_x + \cos k_y) =0$
(in the units where the lattice constant $a=1$). There are four nodes
centered at $(\pm \pi/2, \pm \pi/2)$ around which the energy spectrum is 
linearized. The corresponding nodal sub-zones 
(see Eq.~(\ref{nodal.approximation})) are called BZ $j$ with $j = 1, \ldots,4$.
For $\alpha_{D} \ll 1$
the Landau damping develops only if the direction of the phason
momentum $\mathbf{K}$ is within one of the ``cones''.
The size of these cones is dependent on the ratio $c = \Omega/v_F K$,
so that the cones shrink as $|c| \to 1$ 
(see Eqs.~(\ref{f1}) - (\ref{f3})).}  
\label{fig:1}
\end{figure}

\begin{figure}[h]
\centering{
\includegraphics[width=8cm]{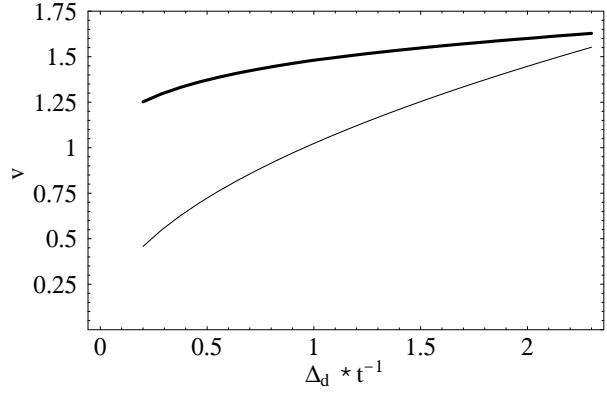}}
\caption{The dependence of the Goldstone mode velocity, $v(T=0)$ 
on the amplitude of the gap, $\Delta_{d}$. Solid line is the result
of numerical calculation with $\xi({\bf k}) = -2 (\cos k_x + \cos k_y)$,
$\Delta({\bf k}) = \Delta_{d}/2 (\cos k_x - \cos k_y)$
(we put $t =a =1$, so that $\Delta_{d}$ is expressed in units of $t$).
Thin line is obtained using Eq.~(\ref{BKTmode.d}). The anisotropy
of the Dirac spectrum for this case is $\alpha_{D} = 4/\Delta_{d}$,
so that $\Delta_{d} =0.2$ corresponds to $\alpha_{D}= 20$.}  
\label{fig:2}
\end{figure}

Using (\ref{nodal.approximation}) one can express Eq.~(\ref{kinetic.1.final})
in terms of $v_{F}$ and $v_{\Delta}$
\begin{equation}\label{kinetic.1.nodal}
\begin{split} 
\Omega_{\mathrm{kin}}^{(1)} & = T \int_{0}^{\beta} d \tau \int d^2 r
\left[ i n_f \frac{\partial_{\tau} \theta}{2} + 
\int_{0}^{p_{max}} \frac{p d p}{2 \pi v_{F} v_{\Delta}} \int_{0}^{2 \pi}
\frac{d \varphi}{2 \pi}  \tanh \frac{p}{2T} \frac{a^2 p
\cos^2 \varphi }{4} (\nabla \theta)^2 \right] \\
& \approx T \int_{0}^{\beta} d \tau \int d^2 r 
\left[i n_f \frac{\partial_{\tau} \theta}{2} + \frac{\sqrt{\pi v_{F}
v_{\Delta}}}{48 a} (\nabla \theta)^2 \right]\,,
\end{split}
\end{equation}
where
\begin{equation}
\label{number.eq}
n_f = \int \frac{d^2 k}{(2 \pi)^2} n({\bf k})
\end{equation}
is the density of carriers.
We note that due to the slow convergence of the integrand in
(\ref{kinetic.1.nodal}) the final expression depends explicitly on the value
of the momentum cutoff $p_{max}$ which was defined in \cite{Lee} in  such a way
that the area of the new integration region over 4 Brillouin sub-zones
(see Fig.~\ref{fig:1}) is the same as that of the original
Brillouin zone. Going from Eq.~(\ref{kinetic.1.final}) to 
(\ref{kinetic.1.nodal}) we essentially replace the averaging over the 
true Fermi surface of the system by the averaging over 4 nodal sub-zones.
The validity of this approximation can only be justified if the
corresponding integrals contain the derivative of the Fermi distribution
$n_{F}({\bf k})$ which is highly peaked in the vicinity of the nodes.
This appears to be the case of the temperature dependent parts
of the phase stiffness $J(T)$, compressibility $K(T)$ and the 
Landau damping terms. For the zero temperature values $J(T =0)$
and $K(T=0)$ the nodal approximation is not well justified.
However, as we show in Sec.~\ref{sec:local.action}, this approximation can
be justified {\it a forteriory} for their ratio (see Fig.~\ref{fig:2})
which determines the velocity of the Bogolyubov-Anderson-Goldstone mode. 
Finally, we stress that after approximation is used, it is impossible to 
recover the $s$-wave limit by putting $v_{\Delta} \to 0$.

\section{The Second Order Term of the Effective Action}
\label{sec:second.tr}

Let us evaluate the trace of the second term
in expansion (\ref{Omega.Kinetic.phase}):
\begin{equation}
\label{kinetic.2}
 \Omega_{\mathrm{kin}}^{(2)} = \frac{T}{2} \mbox{Tr}
[\mathcal{G} \Sigma \mathcal{G} \Sigma]\,.
\end{equation}
Substituting (\ref{Sigma}) into (\ref{kinetic.2}) we obtain that
\begin{equation}\label{kinetic.2.terms}
\Omega_{\rm kin}^{(2)} = \Omega_{\rm kin}^{(2)} \{\theta \Omega_n^2 \theta\}+
\Omega_{\rm kin}^{(2)} \{ \theta \mathbf{K}^2 \theta\}+ 
\Omega_{\rm kin}^{(2)} \{\theta \mathbf{K} \Omega_n \theta \}\,,
\end{equation}
where using $\{\theta \cdots \theta \}$ we denoted symbolically that the
corresponding term of (\ref{kinetic.2}) is either diagonal
(i.e. its frequency-momentum representation contains 
$\theta(i \Omega_n, \mathbf{K}) \Omega_n^2 \theta(- i \Omega_n, -\mathbf{K})$ 
or  $\theta(i \Omega_n, \mathbf{K}) \mathbf{K}^2 
\theta(- i \Omega_n, -\mathbf{K})$) as 
\begin{equation}
\label{kinetic.frequency}
%\begin{split}
\beta \Omega_{\rm kin}^{(2)} \{\theta \Omega_n^2 \theta\} 
 = \frac{T}{2} \sum_{n= -\infty}^{\infty} 
\int \frac{d^2 K}{(2 \pi)^{2}}\, \theta(i\Omega_n, {\bf K})
\left(-\frac{\Omega_n^2}{4} \right) \theta(-i\Omega_n, -{\bf K}) 
 T \sum_{l = - \infty}^{\infty} \int \frac{d^2 k}{(2 \pi)^2}\,
\pi_{33}(i\Omega_{n}, {\bf K}; i \omega_{l}, {\bf k})\,;
%\end{split}
\end{equation}
\begin{equation}
\label{kinetic.momentum}
\begin{split}
\beta & \Omega_{\rm kin}^{(2)} \{ \theta \mathbf{K}^2 \theta \} = 
\frac{T}{2} \sum_{n = -\infty}^{\infty} 
\int \frac{d^2 K}{(2 \pi)^{2}}\, \theta(i\Omega_n, {\bf K})
K_{x}^{2} \theta(-i\Omega_n, -{\bf K}) \times \\
& T \sum_{l = - \infty}^{\infty} \int \frac{d^2 k}{(2 \pi)^2}\, 
\pi_{00}(i\Omega_{n}, {\bf K}; i \omega_{l}, {\bf k}) 
(ta)^2 \sin (k_x - K_x/2)a \sin (k_x + K_x/2)a 
 + (x \to y)\,
\end{split}
\end{equation}
or mixed
\begin{equation} \label{kinetic.frequency.momentum}
\begin{split}
 \beta \Omega_{\rm kin}^{(2)} & \{ \theta \mathbf{K} \Omega_n \theta \} = - T
\sum_{n = -\infty}^{\infty} \int \frac{d^2 K}{(2 \pi)^{2}}\, \theta(i\Omega_n,
{\bf K})
\frac{K_{x} \Omega_n}{2} \theta(-i\Omega_n, -{\bf K}) \\
& \times T \sum_{l = - \infty}^{\infty} \int \frac{d^2 k}{(2 \pi)^2}\,
\left[ \pi_{03}(i\Omega_{n},{\bf K}; i \omega_{l}, {\bf k}) 
\frac{i t a}{2} \sin (k_x + K_x/2)a + 
\pi_{30}(i\Omega_{n},{\bf K}; i \omega_{l}, {\bf k}) 
\frac{i t a}{2} \sin (k_x - K_x/2)a  \right] \\
& + (x \to y)\,.
\end{split}
\end{equation}
In (\ref{kinetic.frequency}) - (\ref{kinetic.frequency.momentum})
we introduced the following short-hand notations
\begin{equation}\label{pi}
\pi_{ij}(i\Omega_{n},{\bf K}; i \omega_{l}, {\bf k}) \equiv
\mbox{tr} [\mathcal{G} (i \omega_{l} + i\Omega_{n}, {\bf k} + {\bf K}/2)
\tau_{i} \mathcal{G} (i \omega_{l}, {\bf k} - {\bf K}/2) \tau_{j}]\,,
\qquad
\tau_{i} = (\tau_{0} \equiv \hat{I}, \tau_{3})\,.
\end{equation}

More generally, we can rewrite Eqs.~(\ref{kinetic.momentum}) 
and (\ref{kinetic.frequency.momentum}) as follows
\begin{equation}\label{kinetic.momentum.general}
\beta  \Omega_{\rm kin}^{(2)} \{ \theta \mathbf{K}^2 \theta \} \simeq
\frac{T}{2} \sum_{n = -\infty}^{\infty} \int \frac{d^2 K}{(2 \pi)^{2}}\,
\theta(i\Omega_n, {\bf K})
\frac{K_{\alpha} K_{\beta}}{4} \theta(-i\Omega_n, -{\bf K}) 
\Pi_{00}^{\alpha \beta}(i\Omega_{n}, {\bf K})\,;
\end{equation}
\begin{equation}
\label{kinetic.frequency.momentum.general}
 \beta \Omega_{\rm kin}^{(2)}  \{ \theta \mathbf{K} \Omega_n \theta \} =  
- \frac{T}{2}
\sum_{n = -\infty}^{\infty} \int \frac{d^2 K}{(2 \pi)^{2}}\, 
\theta(i\Omega_n, {\bf K})
\frac{ i \Omega_n K_{\alpha}}{4} \theta(-i\Omega_n, -{\bf K}) \,
[\Pi_{03}^{\alpha}(i\Omega_{n}, {\bf K}) +
\Pi_{30}^{\alpha}(i\Omega_{n}, {\bf K})]\,,
\end{equation}
where 
\begin{equation}\label{Pi00}
\begin{split}
& \Pi_{00}^{\alpha \beta}(i\Omega_{n}, {\bf K}) \equiv
\sum_{l = - \infty}^{\infty} \int \frac{d^2 k}{(2 \pi)^2}\,
\pi_{00}(i\Omega_{n},{\bf K}; i \omega_{l}, {\bf k})
v_{F \alpha}({\bf k}) v_{F \beta}({\bf k})\,, \\
& \Pi_{03}^{\alpha}(i\Omega_{n}, {\bf K}) \equiv
\sum_{l = - \infty}^{\infty} \int \frac{d^2 k}{(2 \pi)^2}\,
\pi_{03}(i\Omega_{n},{\bf K}; i \omega_{l}, {\bf k})
v_{F \alpha}({\bf k}) \,, \\
\end{split}
\end{equation}
$v_{F \alpha}(\mathbf{k})  = 
\partial \xi(\mathbf{k})/\partial k_{\alpha}$,
we used the approximation $v_{F \alpha}(\mathbf{k}) \simeq
v_{F \alpha}(\mathbf{k} \pm \mathbf{K}/2)$ and took into account that
$(1/2 \pi)\int d^2 k v_\alpha v_{\beta} = 
\int k dk v^2 \delta_{\alpha \beta}/2$.
It is convenient to introduce here
\begin{equation}
\Pi_{33}(i\Omega_{n}, {\bf K}) \equiv
\sum_{l = - \infty}^{\infty} \int \frac{d^2 k}{(2 \pi)^2}\,
\pi_{33}(i\Omega_{n},{\bf K}; i \omega_{l}, {\bf k})\,
\end{equation}
which would allow to rewrite (\ref{kinetic.frequency})
in the same fashion as (\ref{kinetic.momentum.general}).

As one could notice the product of the Fermi velocities enters 
Eq.~(\ref{kinetic.momentum.general}) via (\ref{Pi00}). 
There is nothing surprising in this 
fact since this piece of the effective action is related to the paramagnetic 
current correlator $\langle j_{\alpha} j_{\beta} \rangle$ 
\cite{Randeria.action}
and in its turn the current operator, $\mathbf{j}$ contains the Fermi 
velocity ${\bf v}_{F}({\bf k})$.
We will return to this point considering the Landau
term which originates from (\ref{kinetic.momentum.general}), so that here we
note only that the current correlator term along with
the diamagnetic term $\sim (\nabla \theta)^2$ in Eq.~(\ref{kinetic.1})
form together the mean-field phase stiffness.       

The matrix traces $\pi_{ij}$ and the corresponding expressions
for $\Pi_{ij}$ are calculated in  Appendix~\ref{sec:A}. Although
the expressions for them are rather lengthy they have a clear
physical interpretation which is also discussed in the Appendix.

\section{The effective Lagrangian at $T \neq 0$ without the Landau terms}
\label{sec:local.action}

The contribution of the first order term to the effective action is given in
(\ref{kinetic.1.nodal}). Concerning the second order term, we note that 
when the Landau terms are neglected it is enough to
set $\Omega_n=0$ inside $\Pi$ in (\ref{kinetic.frequency}),
(\ref{kinetic.momentum.general}) and 
(\ref{kinetic.frequency.momentum.general}). To be more precise, 
the Landau terms arise from the second line of Eqs.~(\ref{A}) 
and (\ref{tr.mixed}) which contains ``dangerous'' denominators 
$1/(E_{+} -E_{-} \pm i \Omega_n)$. One can
however notice that the second line of Eq.~(\ref{A}) leads also to the
regular terms which are proportional to the derivative $d n_{F}(E)/dE$. 
These lead to the second term  in the square brackets
in (\ref{kinetic.time}) and the whole expression (\ref{kinetic.space})
shown below.
For the $s$-wave superconductivity \cite{Gusynin.JETP1,Aitchison} in the 
temperature region $0 < T \lesssim 0.6 T_{c}$ these terms are very small 
compared to the main terms. 
Although the contribution from these terms is still local, their
presence breaks the Galilean invariance \cite{Aitchison}.
This is the reason why  it was more natural for \cite{Aitchison} to treat them
along with ``true'' Landau terms since they also originate 
from the same denominators of the second line of (\ref{A}) as was
mentioned above.
For the $d$-wave superconductivity this splitting, however, appears
to be rather artificial since these terms are not small even for low
temperatures due to the presence of the nodal quasiparticles, so here
we will consider all regular terms.

For the regular terms from the second order term  we obtain a local
effective action, involving time and space derivatives of $\theta(t, {\bf r})$:
\begin{equation}
\label{kinetic.time}
\begin{split}
& \beta \Omega_{\rm kin}^{(2)} \{(\partial_{t} \theta)^2\} =  
 \frac{i}{2} \int \frac{d \Omega}{2 \pi} 
\int \frac{d^2 K}{(2 \pi)^{2}}\, \theta(\Omega, {\bf K})
\frac{\Omega^2}{4} \theta(-\Omega, -{\bf K})  
\Pi_{33}(0, {\bf K} \to 0) \\ & =   i
\frac{T}{2} \int d t \int d^2 r \frac{(\partial_t \theta(t, {\bf
r}))^2}{4}\int \frac{d^2 k}{(2 \pi)^2} \left[- \frac{\Delta^2({\bf k})}{E^3({\bf
k})} \tanh \frac{E({\bf k})}{2T} - \frac{1}{2T} \frac{\xi^2({\bf k})}{E^2({\bf
k})} \cosh^{-2} \frac{E({\bf k})}{2T}\right]\,,
\end{split}
\end{equation}
\begin{equation}\label{kinetic.space}
\begin{split}
\beta \Omega_{\rm kin}^{(2)} &\{ (\nabla \theta)^2\} = 
 \frac{i}{2} \int \frac{d \Omega}{2 \pi} \int
\frac{d^2 K}{(2 \pi)^{2}}\, \theta(\Omega, {\bf K})
\frac{K_{\alpha} K_{\beta}}{4} \theta(-\Omega, -{\bf K}) 
\Pi_{00}^{\alpha \beta}(0, {\bf K \to 0}) \\
& = i \frac{T}{2} \int d t \int d^2 r \frac{\nabla_{\alpha} \theta(t, {\bf r})
\nabla_{\beta} \theta(t, {\bf r}) }{4}\int \frac{d^2 k}{(2 \pi)^2} \left[
-\frac{1}{2T} \cosh^{-2} \frac{E({\bf k})}{2T}\right] v_{F \alpha}(\mathbf{k})
v_{F \beta}(\mathbf{k})
\end{split}
\end{equation}
and the mixed term (\ref{kinetic.frequency.momentum.general}) 
does not contribute to the regular part within the used
approximation, since $\Pi_{30}^{\alpha}(0, {\bf K} \to 0) =0$. 
Evaluating (\ref{kinetic.time}) and (\ref{kinetic.space}) 
we performed the analytical continuation $i \Omega_n \to
\Omega + i0$ back to the real continuous frequencies, so that $t$ is the real
time.

Using the nodal expansion (\ref{nodal.approximation}) to calculate
(\ref{kinetic.time}), (\ref{kinetic.space}) and adding (\ref{kinetic.1.nodal})
we finally obtain the regular effective Lagrangian, $\mathcal{L}^{R}$
such that $\beta \Omega_{\rm kin} = - i \int dt \int d^2 r 
\mathcal{L}^{\mbox{\tiny R}}(t, \mathbf{r})$  
for $T \ll \Delta_{d}$ and ignoring the Landau terms:
\begin{equation}
\label{Lagrangian}
\mathcal{L}^{\mbox{\tiny R}} = - 
\frac{n_f}{2} \partial_t \theta(t, {\bf r}) + \frac{K}{2}
 (\partial_{t} \theta(t, {\bf r}))^2 - 
\frac{J}{2}(\nabla \theta (t, {\bf r}))^2\,,
\end{equation}
where the phase stiffness $J \equiv J_{d,s}$ and 
compressibility $K \equiv K_{d,s}$ are
\begin{equation}
\label{kinetic.d-wave}
J_d = \frac{\sqrt{\pi v_{F} v_{\Delta}}}{24 a} - \frac{\ln 2}{2 \pi}
\frac{v_F}{v_{\Delta}}T\,, \qquad  
K_d = \frac{1}{4 a \sqrt{\pi v_{F} v_{\Delta}}}\,.
\end{equation}
The linear time derivative term in
(\ref{Lagrangian}) is important for the description of vortex
dynamics (see \cite{Randeria.action} and Refs. therein), but we 
omit it in what follows.
% We note also that the term $\sim T^3$ from (\ref{kinetic.1.nodal}) 
%was not included in Eq.~(\ref{kinetic.d-wave}) because it has the same order
%as the terms which would appear only if one takes into 
%account that $\Delta_d$ and consequently $v_{\Delta}$ are dependent
%on $T$. 
We stress that the second, temperature dependent term in $J_d$
follows from Eq.~(\ref{kinetic.space}) which contains the derivative
of the Fermi distribution, $dn_{F}(E)/dE$. Its presence, as we mentioned
in Sec.~\ref{sec:first.tr}, makes the nodal approximation valid
\cite{Lee}.
Since we consider only the low temperature
$T \ll \Delta_{d}$ region, we restrict ourselves by the values
$\Delta_d$ and $v_{\Delta}$ at $T =0$, so that all temperature 
dependences appear to be linear. For higher temperatures
it is necessary to take into account that $\Delta_{d}(T)$ and
$v_{\Delta}(T)$ are in fact  decreasing functions of $T$.

It is useful to compare the stiffness and compressibility 
from (\ref{kinetic.d-wave}) with those parameters derived 
in \cite{Gusynin.JETP1} for the continuum 2D model
with $s$-wave pairing:
\begin{equation}
\label{kinetic.s-wave}
 J_s = \frac{n_f}{4m} \left( 1 - \int_{0}^{\infty} d x 
\frac{1}{\cosh^2 \sqrt{x^2 + \dfrac{\Delta_s^2}{4T^2}}} \right)\,,
\qquad K_s = \frac{m}{4 \pi} \,.
\end{equation}

First of all one can see that for low temperatures the superfluid
stiffness in (\ref{kinetic.s-wave}) does not contain any term which 
goes to zero more slowly
than $\exp(- \Delta_s/T)$, while (\ref{kinetic.d-wave})
has a term proportional to $T$. The origin of this difference is
well-known and related to the presence of the nodal quasiparticles.
Secondly, we can compare the values of the zero temperature superfluid 
stiffness which for the continuum translationally invariant system has to 
be equal to $n_f/4m$, so that all carriers participate
in the superfluid ground state \cite{Leggett}. Since the presence of 
a lattice evidently breaks the continuum
translational invariance the superfluid
density at $T=0$ in the case of (\ref{kinetic.d-wave}) is less than 
$n_f /4m$, as can be readily seen from (\ref{kinetic.1.final}).
Finally, since we consider here a neutral system it has the 
Berezinskii-Kosterlitz-Thouless (BKT) collective mode 
\cite{foot2}, $\Omega^2 = v^2 K^2$ with $v = \sqrt{J/K}$. 
One can see that for the continuum $s$-wave case $v = v_{F}/\sqrt{2}$ and
\begin{equation}
\label{BKTmode.d}
v = \sqrt{\frac{\pi v_{F} v_{\Delta}}{6} - 
\frac{2 \ln 2 a v_{F}}{\sqrt{\pi}} \sqrt{\frac{v_{F}}{v_{\Delta}}}T} 
\end{equation}
for the $d$-wave model on lattice at $T \sim 0$.
Eq.~(\ref{BKTmode.d}) gives rather simple approximate expression
for the BKT mode velocity for the lattice model of
$d$-wave superconductor. Comparing in Fig.~\ref{fig:2} the results 
obtained for $v(T=0)$ using Eq.~(\ref{BKTmode.d})  with the numerical
computation  without the nodal approximation, we can see that
that even being very simple Eq.~(\ref{BKTmode.d}) predicts the correct
behavior of $v(T=0)$. 

Following \cite{Randeria.action} we estimate the upper
values for the frequency $\Omega$ in (\ref{kinetic.time}) and the momentum $K$
in (\ref{kinetic.space}). The ``phase-only'' effective action
is appropriate for phase distortions whose energy is smaller than
the condensation energy, $E_{cond} \simeq N(0) \Delta^2/2$, where
$N(0)$ is the density of states. For the $s$-wave case (\ref{kinetic.s-wave})
this leads to the following restrictions: $\Omega < \Delta_s$,
$v_{F} K < \Delta_{s}$ and for the $d$-wave case 
(\ref{kinetic.d-wave}): $\Omega < \Delta_{d} \sqrt[4]{1/\alpha_{D}}$,
$v_{F}K < \Delta_{d}  \sqrt[4]{\alpha_{D}}$,
where $\alpha_{D}$ is the anisotropy of the Dirac spectrum defined
in Introduction.

\section{The imaginary part of the Landau terms for 2D $d$-wave 
and $s$-wave cases}
\label{sec:Landau.damping}

The key values which are necessary for evaluation of the Landau terms are the
differences $E_{+} - E_{-}$ and $n_{F}(E_{+}) -n_{F}(E_{-})$. 
Expanding in $\mathbf{K}$, 
\begin{equation}\label{difference.defintion}
E({\bf k} + {\bf K}/2) - E({\bf k} - {\bf K}/2) = 
\mathbf{v}_{g}(\mathbf{k}) \mathbf{K}\,,
\end{equation}
where the group velocity is given by
\begin{equation}\label{group.velocity}
\mathbf{v}_{g}(\mathbf{k}) = \nabla_{\mathbf{k}} E(\mathbf{k}) =
\frac{1}{E(\mathbf{k})} \left[\xi(\mathbf{k}) \mathbf{v}_{F} + 
\Delta(\mathbf{k}) \mathbf{v}_{\Delta}\right]\,.
\end{equation}
It is obvious that due to the gap $\mathbf{k}$-dependence
Eq.~(\ref{difference.defintion}) differs from the $s$-wave case \cite{Aitchison}, where the difference is simply
\begin{equation}\label{difference.s}
E({\bf k} + {\bf K}/2) - E({\bf k} - {\bf K}/2) = 
\frac{\xi(\mathbf{k})}{E(\mathbf{k})} \mathbf{v}_{F} \mathbf{K} \qquad
(\Delta(\mathbf{k}) = \Delta_{s})\,.
\end{equation}
It is convenient to rewrite (\ref{difference.defintion}) and
(\ref{group.velocity}) in terms of the nodal approximation described after 
Eq.~(\ref{nodal.approximation}). In the vicinity of one of the nodes we have
\iffalse
\begin{equation}
\label{dispersion.expand} \xi({\bf k} + {\bf K}/2) \simeq \left.
\frac{\partial \xi}{\partial {\bf k}} \right|_{{\bf k}_0} ({\bf k} + {\bf K}/2
-{\bf k}_0) = v_F \hat{\mathbf{k}}_1 ({\bf k} + {\bf K}/2 -{\bf k}_0) = v_F
k_1 + v_F \hat{\mathbf{k}}_1{\bf K}/2 = p \cos \varphi + v_F K_1/2
\end{equation}
and
\begin{equation}\label{gap.expand}
\Delta({\bf k} + {\bf K}/2) \simeq \left. \frac{\partial \Delta}{\partial {\bf
k}} \right|_{{\bf k}_0} ({\bf k} + {\bf K}/2 -{\bf k}_0) = v_{\Delta}
\hat{\mathbf{k}}_2 ({\bf k} + {\bf K}/2 -{\bf k}_0) = v_{\Delta} k_2 +
v_{\Delta} \hat{\mathbf{k}}_2{\bf K}/2 = p \sin \varphi + v_{\Delta} K_2/2\,,
\end{equation}
\fi
\begin{equation}\label{difference}
{\bf v}_{g} {\bf K} = v_{F} K_1 \cos \varphi +
v_\Delta K_2 \sin \varphi \equiv P \cos(\varphi - \psi)\,,
\qquad E({\bf k} \pm {\bf K}/2) \ll \Delta_{d}\,,
\end{equation}
where the momentum $\mathbf{K} = (K_1,K_2)$ of $\theta$-particle was also 
expressed in the nodal coordinate system 
$\hat{\mathbf{k}}_{1}$, $\hat{\mathbf{k}}_{2}$, so that
$P^1 \equiv v_{F} K_1  = P \cos \psi$, $P^2 \equiv v_{\Delta} K_2 = P\sin
\psi$ and 
$P = \sqrt{(P^1)^2 + (P^2)^2} = \sqrt{v_{F}^2 K_1^2 + v_\Delta^2 K_2^2}$.
(We denoted the components of ${\bf P}$ as $P^1, P^2$ to
make them different from the node label $P_{j}$ used in what follows.)

The corresponding
substitution in the integrals over $\mathbf{K}$ reads
similarly to Eq.~(\ref{nodal.approximation})
\begin{equation}\label{nodal.approximation.phason}
\int \frac{d^2 K}{(2 \pi)^2} \to  \sum_{j =1}^{4}\int_{0}^{P_{max}} 
\frac{P_j d P_j}{2 \pi v_{F} v_\Delta} \int_{0}^{2 \pi} 
\frac{d \psi_j}{2 \pi}\,,
\end{equation}
where $P_{max}$ is evidently related to the maximal value of
$K$ discussed at the end of Sec.~\ref{sec:local.action}.
Finally, we can approximate the difference $n_{F}(E_{+})- n_{F}(E_{-})$
as
\begin{equation}\label{Fermi.difference}
n_{F}(E_{+})- n_{F}(E_{-}) \approx \frac{d n_{F}(E)}{d E} 
{\bf v}_{g} {\bf K} =  \frac{d n_{F}(E)}{d E} P \cos (\varphi - \psi)
\end{equation}

Having the differences (\ref{difference}) and (\ref{Fermi.difference})
we can now derive the imaginary part for the Landau terms. 
In the subsequent subsections we consider all three 
terms of Eq.~(\ref{kinetic.2.terms}) for $d$-wave pairing
and compare them with their $s$-wave counterparts.
We would like to note that the Landau terms have also
the {\it real part} which for the 3D $s$-wave case was consider
in detail in \cite{Aitchison}. This real part consists of 
regular and irregular terms. The regular term was already taken into 
account in Sec.~\ref{sec:local.action} and the irregular term
is not considered in this paper.

\subsection{$\Omega_{\rm kin}^{(2)} \{\theta \Omega^2 \theta\}$ term}
Let us consider firstly the contribution from
$\Omega_{\rm kin}^{(2)} \{\theta \Omega_n^2 \theta\}$ (see
Eqs.~(\ref{kinetic.frequency}) and (\ref{A}) for $A_{-}$), which
after the analytical continuation $i \Omega_n \to \Omega +i0$ takes the form
\begin{equation}\label{kinetic.frequency.Landau}
\begin{split}
& \mbox{Im}[i \beta \Omega_{\rm kin}^{(2)} \{\theta \Omega^2 \theta\}]  
\approx -
\frac{1}{2} \int \frac{d \Omega d^2 K}{(2 \pi)^3} \theta(\Omega,\mathbf{K})
\frac{\Omega^2}{4} \theta(-\Omega,-\mathbf{K})  \\
& \times\int \frac{d^2 k}{(2 \pi)^2} \frac{1}{2} \left(1 +
\frac{\xi^{2} - \Delta^{2}}{E^{2}} \right) \mbox{Im}
\frac{2}{{\bf v}_{g} {\bf K} + \Omega + i0}  \frac{dn_{F}(E)}{d E}
{\bf v}_{g} {\bf K}  \\
& \approx  \sum_{j =1}^{4}  \int \frac{d \Omega}{2 \pi} \int_{0}^{P_{max}}
\frac{P_j d P_j}{2 \pi v_{F} v_\Delta} \int_{0}^{2 \pi} \frac{d \psi_j}{2 \pi}
\theta(\Omega,\mathbf{K})
\frac{\Omega^2}{8} \theta(-\Omega,-\mathbf{K}) \\
& \times \int_{0}^{p_{max}} \frac{p d p}{2 \pi v_{F} v_\Delta}  \int_{0}^{2\pi}
d \varphi  \cos^2 \varphi  \delta(P_j
\cos(\varphi - \psi_j) + \Omega) \frac{dn_{F}(E)}{d E} P_j \cos(\varphi - \psi_j) \\
& = \sum_{j =1}^{4} \int \frac{d \Omega}{2 \pi} \int_{0}^{P_{max}} \frac{P_j d
P_j}{2 \pi v_{F} v_\Delta} \int_{0}^{2 \pi} \frac{d \psi_j}{2 \pi}
\theta(\Omega,\mathbf{K}) \frac{\Omega^2}{8} \theta(-\Omega,-\mathbf{K})
\frac{\ln 2}{\pi} \frac{T}{v_{F} v_{\Delta}}\\
& \times \frac{1}{\sqrt{1 - \dfrac{\Omega^2}{P_j^2}}}  \frac{\Omega}{P_j}
\left[\frac{\Omega^2}{P_j^2} \cos^2 \psi_j +   
\left(1- \frac{\Omega^2}{P_j^2} \right) \sin^2 \psi_j  \right]
 \Theta \left(1 - \frac{|\Omega|}{P_j} \right)\,.
\end{split}
\end{equation}
Here $\Theta(x)$ is the step function and we used  the integral
\begin{equation}\label{integral}
\int_{0}^{\infty} \frac{p d p}{2 \pi v_{F} v_\Delta} \frac{1}{2T}
\frac{1}{\cosh^2 \dfrac{p}{2T}} = \frac{\ln 2}{\pi} \frac{1}{v_{F} v_{\Delta}}
T\,.
\end{equation}
Integrating over $\varphi$ we had to take into account that there are two
points where $\delta$-function contributes into the integral:
$\cos \varphi = - \Omega/P_j$, 
$\sin \varphi = \sqrt{1 - \Omega^2/P_j^2}$ and 
$\cos \varphi = - \Omega/P_j$, 
$\sin \varphi = -\sqrt{1 - \Omega^2/P_j^2}$. 

As we can see from the first equality in
(\ref{kinetic.frequency.Landau}) the imaginary
part develops when $\Omega = E_{+} - E_{-} \approx {\bf v}_{g} {\bf K}$.
This condition was interpreted in \cite{Abrahams} as the ``{\v C}erenkov''
irradiation (absorption) condition for the process: 
``thermally excited $\theta$-fluctuation quantum (phason) + quasiparticle
$\longleftrightarrow$ phason + quasiparticle'' (or
phason being absorbed and scattering thermally excited quasiparticles).  
As was discussed in \cite{Lee}, since  definite energy
is carried by the quasiparticles, this process is defined by their
group velocity ${\bf v}_{g} = \partial E({\bf k})/\partial {\bf k}$
\cite{Abrahams} as well as thermal and spin currents.

It is in fact a coincidence that for the $s$-wave superconductor the directions
of ${\bf v}_{g}$ and ${\bf v}_{F}$ are the same, so that for the 2D $s$-wave
superconductor using (\ref{difference.s}) instead of 
(\ref{difference.defintion}) and (\ref{group.velocity})
one can obtain 
\begin{equation}\label{kinetic.frequency.Landau.s}
\begin{split}
& \mbox{Im}[i \beta \Omega_{\rm kin}^{(2)} \{\theta \Omega^2 \theta\}]  = -
\frac{1}{2} \int \frac{d \Omega d^2 K}{(2 \pi)^3} \theta(\Omega,\mathbf{K})
\frac{\Omega^2}{4} \theta(-\Omega,-\mathbf{K})  
\int \frac{d^2 k}{(2 \pi)^2} \frac{2 \xi^2}{E^2} \mbox{Im}
\frac{1}{\dfrac{\xi}{E} {\bf v}_{F} {\bf K} + \Omega + i0} \frac{dn(E)}{dE}
\frac{\xi}{E} {\bf v}_{F} {\bf K} \\
&=  - 
\frac{1}{2} \int \frac{d \Omega d^2 K}{(2 \pi)^3} \theta(\Omega,\mathbf{K})
\frac{\Omega^2}{4} \theta(-\Omega,-\mathbf{K})
\frac{\Delta_s m c}{2 \pi} \int_{-\infty}^{\infty} \frac{dy}
{\sqrt{1 - c^2 \dfrac{1+y^2}{y^2}}} \frac{dn}{dE} \frac{2 |y|}{\sqrt{1+y^2}}
\Theta\left(\frac{|y|}{\sqrt{1+y^2}} - |c| \right)\,, 
\end{split}
\end{equation}
where following \cite{Aitchison} we used the notations
$y = \xi/\Delta_s$ and $c = \Omega / v_F K$.
Comparing (\ref{kinetic.frequency.Landau.s}) with the 3D case \cite{Aitchison}
(in the notations of \cite{Aitchison} the term we consider
is related to the sum ${\tilde B}_{L}+ {\tilde C}_{L}$)
one can notice that the only difference between the corresponding
expressions is in the square root in the denominator of  
(\ref{kinetic.frequency.Landau.s}). It obviously originates from
the different measures for the angular integration in 2D and 3D where
the extra $\sin \varphi$ is present. Comparing also 
(\ref{kinetic.frequency.Landau}) and (\ref{kinetic.frequency.Landau.s})
one can see that their main analytical structure appears to be
the same, viz. $\sim \Omega^3/P_{j}$ for the $d$-wave case 
and $\sim \Omega^3/v_{F}K$ for the $s$-wave case, so that the momentum 
variables $P_j$ and $v_F K$ do not appear in the numerator.

We  note that the calculation of the ultrasonic attenuation
$\alpha_{\rm sound}(T, \mathbf{K})$ in $d$-wave superconductors results in the
expression
\begin{equation}
\label{ultrasonic}
\alpha_{\rm sound}(T, \mathbf{K}) \sim 
\frac{\Omega}{2T} \int \frac{d^2 k}{(2 \pi)^2}
\frac{\xi^2(\mathbf{k})}{E^2(\mathbf{k})} 
\frac{1}{\cosh^2 \dfrac{E(\mathbf{k})}{2T}} \delta(\mathbf{v}_g \mathbf{K})\,,
\end{equation}
which has the same structure as (\ref{kinetic.frequency.Landau}).
This can be easily seen if one takes into account that
for ultrasound frequency range
$\Omega \ll v_{F} K, \Delta_{d}$, so that
the corresponding terms from Eq.~(\ref{kinetic.frequency.Landau})
can be simplified as follows
\begin{equation}
\label{simplify}
\delta(E_{+} - E_{-} + \Omega) [n_F(E_{-}) - n_F(E_{+})] =
\delta(E_{+} - E_{-} + \Omega) [n_F(\Omega + E_{+}) - n_F(E_{+})]
\simeq \Omega \delta(E_{+} - E_{-}) \frac{n_F(E)}{dE}  
\end{equation} 
leading to Eq.~(\ref{ultrasonic}). This the reason why in the limit
$\Omega/v_{F} K \to 0$ the angular dependence of the Landau damping
(\ref{kinetic.frequency.Landau}) which we consider in 
Sec.~\ref{sec:approximate} would appear to be the same as 
the angular dependence of the ultrasonic attenuation \cite{Carbotte}. 

\subsection{$\Omega_{\rm kin}^{(2)} \{\theta \mathbf{K}^2 \theta\}$ term}
The contribution from $\Omega_{\rm kin}^{(2)} \{\theta \mathbf{K}^2 \theta\}$
(see Eqs.~(\ref{kinetic.momentum.general}) and (\ref{A}) for $A_{-}$) 
can be treated in the same manner
\begin{equation}\label{kinetic.momentum.Landau}
\begin{split}
& \mbox{Im}[i \beta \Omega_{\rm kin}^{(2)} \{\theta \mathbf{K}^2 \theta\}]  
\approx  -
\frac{1}{2} \int \frac{d \Omega d^2 K}{(2 \pi)^3} \theta(\Omega,\mathbf{K})
\frac{K_{\alpha} K_{\beta}}{4} \theta(-\Omega,-\mathbf{K})  \\
& \times\int \frac{d^2 k}{(2 \pi)^2} \frac{1}{2} \left(1 +
\frac{\xi^2 + \Delta^2}{E^2} \right) \mbox{Im}
\frac{2}{\mathbf{v}_{g} \mathbf{K} + \Omega + i0} v_{F \alpha}(\mathbf{k})
v_{F \beta}(\mathbf{k})
\frac{dn_F(E)}{d E} \mathbf{v}_{g} \mathbf{K}\\
& \approx  \sum_{j =1}^{4} \int \frac{d \Omega}{2 \pi} \int_{0}^{P_{max}}
\frac{P_j d P_j}{2 \pi v_{F} v_\Delta} \int_{0}^{2 \pi} \frac{d \psi_j}{2 \pi}
\theta(\Omega,\mathbf{K})
\frac{P_j^2 \cos^2 \psi_j}{8} \theta(-\Omega,-\mathbf{K}) \\
& \times \int_{0}^{p_{max}} \frac{p d p}{2 \pi v_{F} v_\Delta}  \int_{0}^{2\pi}
d \varphi   \delta(P_j
\cos(\varphi - \psi_j) + \Omega) \frac{dn_{F}(E)}{d E} P_j \cos(\varphi - \psi_j)\\
& = \sum_{j =1}^{4} \int \frac{d \Omega}{2 \pi} \int_{0}^{P_{max}} \frac{P_j d
P_j}{2 \pi v_{F} v_\Delta} \int_{0}^{2 \pi} \frac{d \psi_j}{2 \pi}
\theta(\Omega,\mathbf{K}) \frac{P_j^2 \cos^2 \psi_j}{8}
\theta(-\Omega,-\mathbf{K}) \frac{\ln 2}{\pi} \frac{T}{v_{F} v_{\Delta}} \\
& \times \frac{1}{\sqrt{1 - \dfrac{\Omega^2}{P_j^2}}} \frac{\Omega}{P_j} \Theta
\left(1 - \frac{|\Omega|}{P_j} \right)\,
\end{split}
\end{equation}
and for the $s$-wave case
\begin{equation}\label{kinetic.momentum.Landau.s}
\begin{split}
& \mbox{Im}[i \beta \Omega_{\rm kin}^{(2)} \{\theta \mathbf{K}^2 \theta\}]  = -
\frac{1}{2} \int \frac{d \Omega d^2 K}{(2 \pi)^3} \theta(\Omega,\mathbf{K})
\frac{K_{\alpha} K_{\beta}}{4} \theta(-\Omega,-\mathbf{K})  
\int \frac{d^2 k}{(2 \pi)^2}  \mbox{Im}
\frac{2}{\dfrac{\xi}{E} {\bf v}_{F} {\bf K} + \Omega + i0} 
v_{F \alpha} v_{F \beta} \frac{dn(E)}{dE}
\frac{\xi}{E} {\bf v}_{F} {\bf K} \\
&=  - 
\frac{1}{2} \int \frac{d \Omega d^2 K}{(2 \pi)^3} \theta(\Omega,\mathbf{K})
\frac{\Omega^2}{4} \theta(-\Omega,-\mathbf{K})
\frac{\Delta_s m c}{2 \pi} \int_{-\infty}^{\infty} \frac{dy}
{\sqrt{1 - c^2 \dfrac{1+y^2}{y^2}}} \frac{dn}{dE} \frac{2(1+y^2)^{3/2}}{y^2 |y|}
\Theta\left(\frac{|y|}{\sqrt{1+y^2}} - |c| \right)\,. 
\end{split}
\end{equation}
Let us now compare the expressions (\ref{kinetic.momentum.Landau}) and
(\ref{kinetic.momentum.Landau.s}). Evidently their analytical structure
is different since for the $s$-wave case we have again that it is
proportional to $\Omega^3/v_F K$, while for $d$-wave pairing
$P_j$ enters the numerator, so that the main structure
(excluding the measure of integration over $P_j$ and the square
root in the denominator) is $\sim \Omega P_j$. 
This difference originates from the angular integration in 
(\ref{kinetic.momentum.Landau}) and (\ref{kinetic.momentum.Landau.s})
which is performed using the $\delta$-function. Since the arguments
of the $\delta$-function for the $d$- and $s$-wave cases
are different we have obtained two different answers.
Indeed, for the $s$-wave case the argument is proportional to
$\mathbf{v}_{F} \mathbf{K}$ that coincides with the product  
$(\mathbf{v}_{F} \mathbf{K})^2 \dfrac{d n_F(E)}{dE} \mathbf{v}_{F} \mathbf{K}$ 
outside the $\delta$-function.
(see Sec.~\ref{sec:second.tr} after Eq.~(\ref{kinetic.momentum.general}), 
where the origin of the term $\sim (\mathbf{v}_{F} \mathbf{K})^2$
is discussed), so that the angular integration
removes $K$ from the numerator. The $d$-wave case appears to be different
since we have $\mathbf{v}_g \mathbf{K}$ inside the $\delta$-function
and $(\mathbf{v}_{F} \mathbf{K})^2 \dfrac{d n_F(E)}{dE} \mathbf{v}_{g} \mathbf{K}$ outside, so that the angular integration leaves $P_j$ in the numerator.
Despite the fact that $\Omega P_{j}$ looks substantially less
singular than $\Omega/P_j$, the first expression still remains
nonanalytical near $P_j \approx 0$ because it is expressed in terms
of $\sqrt{\mathbf{K}^2}$ the coordinate representation of which is the 
non-local operator $\sqrt{-\nabla^2}$.  

Thus physically the difference between the analytical form of the Landau 
damping in $d$- and $s$-wave superconductors originates from
the $\mathbf{k}$-dependence of the gap  $\Delta_d(\mathbf{k})$
which in its turn makes the direction of the quasiparticle group velocity 
$\mathbf{v}_g$ different from the Fermi velocity $\mathbf{v}_F$.
The last velocity, however, still enters the numerator of  
(\ref{kinetic.momentum.Landau}) since, as was already mentioned in
Sec.~\ref{sec:second.tr}, it originates from the current-current
correlator. The electrical current is proportional to
$\mathbf{v}_F$, not $\mathbf{v}_g$ because quasiparticles carry definite 
energy and spin, but do not carry definite charge.
Therefore, the specific form of the Landau damping terms
in a $d$-wave superconductor has the same physical origin as the source
of the extra terms in the thermal and spin conductivities \cite{Lee}
which are related to the presence in $\mathbf{v}_g$ both  $\mathbf{v}_F$
and  $\mathbf{v}_{\Delta}$ components.

The $s$-wave Landau damping (\ref{kinetic.momentum.Landau.s}) itself
can be again related to the 3D case \cite{Aitchison}, where
it is expressed via the difference $\tilde{B}_{L} - \tilde{C}_{L}$.

\subsection{$\Omega_{\rm kin}^{(2)}\{\theta \mathbf{K} \Omega \theta\}$
term}

Finally we consider the contribution from 
$\Omega_{\rm kin}^{(2)} \{\theta \mathbf{K} \Omega \theta\}$
(see Eqs.~(\ref{kinetic.frequency.momentum.general}) and (\ref{tr.mixed})): 
\begin{equation}\label{kinetic.frequency.momentum.Landau}
\begin{split}
& \mbox{Im}[i \beta \Omega_{\rm kin}^{(2)} \{\theta \mathbf{K} \Omega \theta\}] 
 \approx  -
\frac{1}{2} \int \frac{d \Omega d^2 K}{(2 \pi)^3} \theta(\Omega,\mathbf{K})
\Omega K_{\alpha} \theta(-\Omega,-\mathbf{K})  \\
& \times \int \frac{d^2 k}{(2 \pi)^2} \frac{\xi}{E} 
\mbox{Im}
\frac{1}{\mathbf{v}_{g} \mathbf{K} + \Omega + i0} v_{F \alpha}(\mathbf{k})
\frac{dn_F(E)}{d E} \mathbf{v}_{g} \mathbf{K}\\
& \approx  \sum_{j =1}^{4} \int \frac{d \Omega}{2 \pi} \int_{0}^{P_{max}}
\frac{P_j d P_j}{2 \pi v_{F} v_\Delta} \int_{0}^{2 \pi} \frac{d \psi_j}{2 \pi}
\theta(\Omega,\mathbf{K})
\frac{\Omega P_j}{4} \cos \psi_j \theta(-\Omega,-\mathbf{K}) \\
& \times \int_{0}^{p_{max}} \frac{p d p}{2 \pi v_{F} v_\Delta}  
\int_{0}^{2\pi} d \varphi \cos \varphi  \delta(P_j
\cos(\varphi - \psi_j) + \Omega) \frac{dn_{F}(E)}{d E} P_j \cos(\varphi - \psi_j)\\
& = - \sum_{j =1}^{4} \int \frac{d \Omega}{2 \pi} \int_{0}^{P_{max}} \frac{P_j d
P_j}{2 \pi v_{F} v_\Delta} \int_{0}^{2 \pi} \frac{d \psi_j}{2 \pi}
\theta(\Omega,\mathbf{K}) \frac{\Omega^2}{8}
\theta(-\Omega,-\mathbf{K}) \frac{\ln 2}{\pi} \frac{T}{v_{F} v_{\Delta}} \\
& \times \frac{2}{\sqrt{1 - \dfrac{\Omega^2}{P_j^2}}}  
\frac{\Omega}{P_j} \cos^2 \psi_j 
\Theta \left(1 - \frac{|\Omega|}{P_j} \right)\,
\end{split}
\end{equation}
and for the $s$-wave case
\begin{equation}\label{kinetic.frequency.momentum.Landau.s}
\begin{split}
& \mbox{Im}[i \beta \Omega_{\rm kin}^{(2)} \{\theta \mathbf{K} \Omega \theta\}]  = -
\frac{1}{2} \int \frac{d \Omega d^2 K}{(2 \pi)^3} \theta(\Omega,\mathbf{K})
\Omega K_{\alpha}  \theta(-\Omega,-\mathbf{K})  
\int \frac{d^2 k}{(2 \pi)^2} \frac{\xi}{E} \mbox{Im}
\frac{1}{\dfrac{\xi}{E} {\bf v}_{F} {\bf K} + \Omega + i0} 
v_{F \alpha} \frac{dn(E)}{dE}
\frac{\xi}{E} {\bf v}_{F} {\bf K} \\
&=  
\frac{1}{2} \int \frac{d \Omega d^2 K}{(2 \pi)^3} \theta(\Omega,\mathbf{K})
\Omega^2 \theta(-\Omega,-\mathbf{K})
\frac{\Delta_s m c}{2 \pi} \int_{-\infty}^{\infty} \frac{dy}
{\sqrt{1 - c^2 \dfrac{1+y^2}{y^2}}} \frac{dn}{dE} \frac{(1+y^2)^{1/2}}{|y|}
\Theta\left(\frac{|y|}{\sqrt{1+y^2}} - |c| \right)\,. 
\end{split}
\end{equation}
Once again the difference between $d$- and $s$-wave cases is related to the 
different directions of $\mathbf{v}_g$ and $\mathbf{v}_F$. Since this
time we had  
$\cos \phi \mathbf{v}_{F} \mathbf{K} \dfrac{d n_F(E)}{dE} 
\mathbf{v}_{g} \mathbf{K}$ outside the $\delta$-function, we 
could not get $\Omega P_j$ as in 
Eq.~(\ref{kinetic.momentum.Landau}) and obtained again
$\sim \Omega^3/P_j$ as in Eq.~(\ref{kinetic.frequency.Landau}).
Furthermore, because we have $\cos \varphi$ instead
of $\cos^2 \varphi$ as in  Eq.~(\ref{kinetic.frequency.Landau})
we got only got only $\Omega/P_j \cos \psi_j$ 
(compare with the square brackets in Eq.~(\ref{kinetic.frequency.Landau}))
which after multiplying by  $\Omega^2 P_j$ gives the final
result $\sim \Omega^3/P_{j}$. It is interesting to note that the mixed 
term for $d$- (see Eq.(\ref{kinetic.frequency.momentum.Landau}))
and $s$-wave  (see Eq.(\ref{kinetic.frequency.momentum.Landau.s}))
has the opposite with respect to other terms sign. This means
that the mixed term describes the energy transfer from the 
quasiparticles to the phase excitations. It turns out, however,
that the whole sum (\ref{final.Landau.s}) of the Landau damping terms
has the sign which corresponds to the phason damping.
(We stress that the different sign in front of all final $s$-wave
expressions and their $d$-wave counterparts are due to the explicit
presence of the derivative of the Fermi distribution, which is negative,
in the former expression.)

\subsection{Final expression for the Landau damping term in the $s$-wave case}

Since the Landau damping terms for $s$-wave pairing all are
$\sim \Omega^3/v_F K$, we can combine (\ref{kinetic.frequency.Landau.s}),
(\ref{kinetic.momentum.Landau.s}) and
(\ref{kinetic.frequency.momentum.Landau.s}):
\begin{equation}\label{final.Landau.s}
\begin{split}
&\mbox{Im}[i \beta \Omega_{\rm kin}^{(2)}]  = -
\frac{1}{2} \int \frac{d \Omega d^2 K}{(2 \pi)^3} \theta(\Omega,\mathbf{K})
\Omega^2 \theta(-\Omega,-\mathbf{K}) \\
& \times \frac{\Delta_s m c}{2 \pi} \int_{-\infty}^{\infty} \frac{dy}
{\sqrt{1 - c^2 \dfrac{1+y^2}{y^2}}} \frac{dn}{dE} 
\frac{1}{2 y^2 |y| \sqrt{1 + y^2}}
\Theta\left(\frac{|y|}{\sqrt{1+y^2}} - |c| \right)\,. 
\end{split}
\end{equation}
This expression coincides with the function $H_i^{(1)}(c)$
introduced in \cite{Aitchison} except for the above mentioned 
difference between angular integration in 2D and 3D.
Thus the simultaneous derivation of the $s$- and $d$-wave
cases allowed us to be sure that all relevant terms
were included and to see explicitly why the corresponding
$s$- and $d$-wave terms behave so differently.

\subsection{Temperature and energy-momenta dependences of Landau Damping}
\label{sec:restrictions}

\begin{figure}[h]
\centering{
\includegraphics[width=8cm]{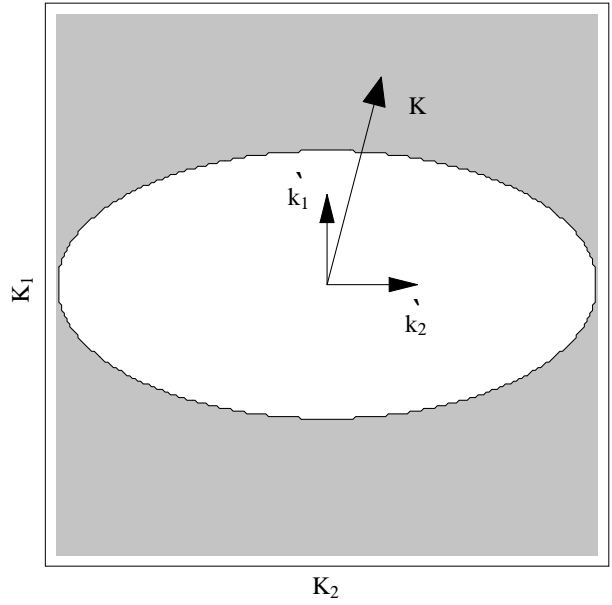}}
\caption{The directional dependence of the Landau damping 
for one node imposed by 
$\Theta\left(1 - 
\frac{|\Omega|}{\sqrt{v_F^{2} K_1^2 + v_{\Delta}^2 K_2^2}}\right)$ for 
$\alpha_{D} =2$. For a given $\Omega$ only excitations with the momenta 
${\bf K}$ which are outside the ellipse (in the shaded region) can 
contribute into the Landau damping. 
For a fixed ratio $|\Omega|/v_{F} |{\bf K}|$  the presence
of the Landau damping depends not only on $|{\bf K}|$, but also
on its direction. For example, for a vector having the length
of the vector ${\bf K}$ shown in the figure the Landau damping 
is possible only if its direction is sufficiently close to
the direction of $\hat{\bf k}_{1}$ normal to the Fermi surface.}  
\label{fig:3}
\end{figure}

We now compare the conditions on the momenta and energies for
the existence of the nonzero Landau terms in $d$- and $s$-wave
cases and the temperature dependences of these terms.
As one can see from Eqs.~(\ref{kinetic.frequency.Landau}),
(\ref{kinetic.momentum.Landau}) and (\ref{kinetic.frequency.momentum.Landau})
for a given node the imaginary parts
develop when $P_j > |\Omega|$ 
(or $\sqrt{v_F^2 K_1^2 + v_{\Delta}^{2} K_{2}^{2}} > |\Omega|$).
Thus in contrast to the $s$-wave case when the imaginary part is nonzero
for all directions of the phason momentum $\mathbf{K}$ satisfying the 
condition $-v_{F} |\mathbf{K}| < \Omega < v_{F} |\mathbf{K}|$
(or $|c| < 1$, where $c$ is defined after 
Eq.~(\ref{kinetic.frequency.Landau.s})), for the $d$-wave it is
sensitive to the direction of the phason momenta as shown in Fig.~\ref{fig:3}.
As we will discuss in Sec.~\ref{sec:approximate} for HTSC 
$v_{F} \gg v_{\Delta}$, so that the directional anisotropy of the Landau
damping becomes very strong. Indeed as can be seen from Fig.~\ref{fig:3}
for $|\Omega|/v_{F} K \lesssim 1$ and $\alpha_{D} \gg 1$ the Landau
damping would exist only in a narrow region of the momenta directions.
Furthermore, if the projection $K_1$ of the phason momentum ${\bf K}$
is not exactly zero (${\bf K} \nparallel \hat{\bf k}_{2}$), 
the condition $|\Omega| < P_{j}$ can be well
replaced by $|\Omega| < v_{F} |K_1|$.
We will discuss the validity of this approximation in 
Sec.~\ref{sec:approximate}.
It is important to stress that the sharp directional
dependence  discussed here is not related to the nodal approximation and
follows only  from the gap anisotropy and the fact that the difference
of the Fermi distribution functions (\ref{Fermi.difference})
which is present in Eqs.~(\ref{kinetic.frequency.Landau}),
(\ref{kinetic.momentum.Landau}) and (\ref{kinetic.frequency.momentum.Landau})
has a very sharp ${\bf k}$-dependence, so that in principle
unbounded (for example, $v_{F}({\bf k}) = {\bf k}/m$ for the model 
with the quadratic dispersion law \cite{Aitchison}) 
functions $v_{g}(\bf{k})$ and $v_{F}({\bf k})$ in 
the corresponding integrals were replaced by their values at the 
Fermi surface.

The final expressions  (\ref{kinetic.frequency.Landau}),
(\ref{kinetic.momentum.Landau}) and (\ref{kinetic.frequency.momentum.Landau})
for the $d$-wave case are in fact
simpler than their $s$-wave counterparts, (\ref{kinetic.frequency.Landau.s}),
(\ref{kinetic.momentum.Landau.s}) and 
(\ref{kinetic.frequency.momentum.Landau.s}) (or the final expression
(\ref{final.Landau.s}))
because the latter expressions
still have one integration which depends on the ratio $|c|$
via the $\Theta$-function. Thus if $|c| = |\Omega|/v_F K \to 1$, only
large $y = \xi/\Delta_s$ contribute into the corresponding integrals.
This circumstance
and the opening of an isotropic gap $\Delta_s$ 
make the contribution from the Landau terms very small comparing to the
main terms. For $d$-wave pairing a relative contribution of
the Landau terms does not depend on the ratio $\Omega/P_j$
via $p$ integration,
so that all frequencies and momenta satisfying the conditions
imposed by the $\Theta$-functions in (\ref{kinetic.frequency.Landau})
and (\ref{kinetic.momentum.Landau}) have the same temperature dependent
weight.

As in the case with the superfluid density the temperature dependence
of the Landau terms is linear in $T$, but while the superfluid
density is nonzero at $T =0$, there is no Landau damping at $T =0$.
The linear $T$ dependence and the absence of damping at $T =0$
are obviously related to the fact that for $d$-wave
pairing there are only four gapless points on the Fermi surface.
It is known, for example, that for a normal (nonsuperconducting) system 
the Lindhard  function (or polarization bubble)
\begin{equation}
\label{Lindhard}
L(\Omega, \mathbf{K}) = \int \frac{d^2 k}{(2 \pi)^2} 
\frac{n_{F}(\xi_{+}) - n_{F}(\xi_{-})}{\xi_{+} - \xi_{-} + \Omega + i0}\,,
\end{equation}
has a nonzero imaginary part even at $T =0$ \cite{Schrieffer} if
the Fermi surface remains ungapped since the derivative of the Fermi
distribution at $T \to 0$ becomes singular on the entire 
Fermi surface (or line in 2D).

Finally, we note that there is also imaginary contribution
even from the first, ``superfluid'' term in (\ref{A}) for $A_{-}$.
As we already mentioned, this term involves pair breaking
which for $s$-wave pairing has the threshold energy, $2 \Delta_s$.
For $d$-wave pairing this energy is zero for $\mathbf{K}=0$
due to the presence of nodes. Nevertheless, if $\mathbf{K} \neq 0$ a
finite energy $P_j$ is necessary to create these excitations
and the imaginary contribution is nonzero only for $|\Omega| > P_j$.
Besides the analytical structure of this term is regular and
it has a higher order than the terms considered here.
Thus this term appears to be less important than
the Landau terms we just considered.

\section{The Approximate Form of the Effective Action and 
$\theta$-propagator}
\label{sec:approximate}

Whereas the local nodal coordinate systems $(P_j, \psi_j)$ are convenient to
write down the corresponding contributions from each node, the final result
should be presented in the global or laboratory
coordinate system $(K, \phi)$. It is
convenient to measure the angle $\phi$ from the vector $\hat{k}_x$,
so that $\phi =0$ corresponds to the corner of the Fermi surface
(see Fig.~\ref{fig:1}) and the first node is at $\phi = \pi/4$. Thus the
transformations from the global coordinate system into the local system
related to the $j$-th node are
\begin{equation}\label{transform}
\begin{split}
& P_j = K \sqrt{v_F^2 \cos^2 \left(\phi -\frac{\pi}{4}+ 
\frac{\pi}{2}(j-1)\right) +
v_{\Delta}^2 \sin^2 \left(\phi - \frac{\pi}{4}+ \frac{\pi}{2}(j-1) \right)}\,, \\
& \cos \psi_j = \frac{v_F K}{P_j} \cos \left(\phi - \frac{\pi}{4}+
\frac{\pi}{2}(j-1)\right)\,, \qquad 
\sin \psi_j = \frac{v_{\Delta} K}{P_j} \sin
\left(\phi - \frac{\pi}{4} + \frac{\pi}{2}(j-1)\right)\,, 
\qquad j =1, \ldots, 4.
\end{split}
\end{equation}

The estimates of \cite{Chiao} show that in YBCO 
$\alpha_{D} \equiv v_{F}/v_{\Delta} =14$
and in BSCCO $\alpha_{D} = 19$ (see  Table~\ref{tab1}).
Thus we can also use the inequality $\alpha_{D} \gg 1$
in what follows. First of all this inequality implies that for 
$j$-th node we may assume that 
\begin{equation}\label{P.approximate}
P_j \approx K v_F \left|
\cos \left(\phi -\frac{\pi}{4}+ \frac{\pi}{2}(j-1)\right)  \right|\,.
\end{equation} 
This approximation is, of course, valid only if $\bf K$ is not
parallel to the Fermi surface (${\bf K} \nparallel \hat{\bf k}_2$).
We note that this direction which is ``dangerous'' for $j$-th node
is the nodal direction for the neighboring nodes. The size
of the ``dangerous'' direction where Eq.~(\ref{P.approximate}) becomes 
invalid can be estimated from the condition 
$K v_{F} \sin \Delta \phi \approx K v_{\Delta} \cos \Delta \phi$ 
which gives $\Delta \phi \approx \alpha_{D}^{-1}$.
Since $\alpha_{D} > 15$ Eq.~(\ref{P.approximate}) is in fact well
justified outside the nodal regions. 

Using  Eq.~(\ref{P.approximate})
we can also rewrite the discussed in Sec.~\ref{sec:restrictions}
inequality $|\Omega| < P_{j}$ imposed by $\Theta$-functions which 
define the region where the {\v C}erenkov condition can be satisfied
as
\begin{equation}\label{condition.approximate}
|\Omega| < v_F K \left|
\cos \left(\phi -\frac{\pi}{4}+ \frac{\pi}{2}(j-1)\right)  \right|\,.
\end{equation}
One can easily see that the condition (\ref{condition.approximate}) 
is in fact equivalent to the condition $|\Omega| < v_F |K_1|$
we already mentioned.

The effective action (\ref{Omega.Kinetic.phase}) in the momentum 
representation in the global coordinate system can be written as
\begin{equation}
\label{effective.action.global}
\beta \Omega_{kin} = 
\frac{i}{2} \int \frac{d \Omega}{2 \pi} \int \frac{K dK}{2 \pi}
\int_{0}^{2 \pi} \frac{d \phi}{2 \pi} \theta(\Omega, \mathbf{K})
[F^{\mbox{\tiny R}}(\Omega, K) + F^{\mbox{\tiny L}}(\Omega, K, \phi)] 
\theta(-\Omega, -\mathbf{K})
\end{equation}
with the regular (compare with Eqs.~(\ref{Lagrangian}) and 
(\ref{kinetic.d-wave}))
\begin{equation}
\label{effective.action.regular}
F^{\mbox{\tiny R}}(\Omega, K) = \frac{K^2}{4}\left( 
\frac{\sqrt{\pi v_{F} v_{\Delta}}}{6 a} - \frac{2\ln 2}{\pi}
\frac{v_F}{v_{\Delta}} T \right) - \frac{\Omega^2}{4} 
\frac{1}{a\sqrt{\pi v_{F} v_{\Delta}}}
\end{equation}
and Landau, $F^{\mbox{\tiny L}}(\Omega, K, \phi)$, parts. 

It is possible to 
write down  $F^{\mbox{\tiny L}}(\Omega, K, \phi)$ substituting 
(\ref{transform}) directly into (\ref{kinetic.frequency.Landau})
and (\ref{kinetic.momentum.Landau}), but to have a more
transparent expression we would like to consider a more simple case
$|\Omega|/v_F K \ll 1$. 
This condition becomes equivalent to $|\Omega|/P_j \ll 1$ due to the 
presence of $\Theta$-functions in (\ref{kinetic.frequency.Landau}), 
(\ref{kinetic.momentum.Landau}) and (\ref{kinetic.frequency.momentum.Landau})
which are cutting out the forbidden domains with $|\Omega|/P_j >1$. 
Physically, the condition $\Omega/v_F K \ll 1$ is
relevant if one, for instance, estimates the Landau damping for
the BKT mode (\ref{BKTmode.d})
because it follows from (\ref{BKTmode.d}) that
$|\Omega|/v_{F} K = \sqrt{\pi/(6 \alpha_{D})} \ll 1$.
Thus we obtain
\begin{equation}
\label{effective.action.singular}
F^{\mbox{\tiny L}}(\Omega, K, \phi) = -i \frac{\Omega^3}{K} \frac{\ln 2}{2 \pi}
\frac{T}{v_{\Delta }v_{F}^{2}} \left[
f_1 \left(\frac{\Omega}{v_F K}, \phi \right) + 
2 f_3 \left(\frac{\Omega}{v_F K}, \phi \right) \right]
-i \Omega K  \frac{\ln 2}{2\pi} \frac{T}{v_{\Delta}}
 f_2 \left(\frac{\Omega}{v_F K}, \phi \right)\,,
\end{equation}
where
\begin{equation}\label{f1}
\begin{split}
f_1 \left(\frac{\Omega}{v_F K}, \phi \right) =
& \frac{\alpha_{D}^{-2} \sin^{2}(\phi -\pi/4)}{[\cos^{2}(\phi -\pi/4) + 
\alpha_{D}^{-2} \sin^{2}(\phi -\pi/4)]^{3/2}} 
\Theta \left(|\cos (\phi - \pi/4)| - \frac{|\Omega|}{v_F K} \right)\\
 + &
\frac{\alpha_{D}^{-2}\cos^{2}(\phi -\pi/4)}{[\sin^{2}(\phi -\pi/4) +
\alpha_{D}^{-2} \cos^{2}(\phi -\pi/4)]^{3/2}} 
\Theta \left(|\sin (\phi - \pi/4)| - \frac{|\Omega|}{v_F K} \right)\,,
\end{split}
\end{equation}
\begin{equation}\label{f2}
\begin{split}
f_2 \left(\frac{\Omega}{v_F K}, \phi \right) =
& \frac{\cos^{2}(\phi -\pi/4)}{[\cos^{2}(\phi -\pi/4) +
\alpha_{D}^{-2} \sin^{2}(\phi -\pi/4)]^{1/2}} 
\Theta \left(|\cos (\phi - \pi/4)| - \frac{|\Omega|}{v_F K} \right)\\
 + &
\frac{\sin^{2}(\phi -\pi/4)}{[\sin^{2}(\phi -\pi/4) +
\alpha_{D}^{-2} \cos^{2}(\phi -\pi/4)]^{1/2}} 
\Theta \left(|\sin (\phi - \pi/4)| - \frac{|\Omega|}{v_F K} \right)
\end{split}
\end{equation}
and 
\begin{equation}\label{f3}
\begin{split}
f_3 \left(\frac{\Omega}{v_F K}, \phi \right) =
& - \frac{\cos^2 (\phi -\pi/4)}{[\cos^{2}(\phi -\pi/4) + 
\alpha_{D}^{-2} \sin^{2}(\phi -\pi/4)]^{3/2}} 
\Theta \left(|\cos (\phi - \pi/4)| - \frac{|\Omega|}{v_F K} \right)\\
& - 
\frac{\sin^2 (\phi -\pi/4)}{[\sin^{2}(\phi -\pi/4) +
\alpha_{D}^{-2} \cos^{2}(\phi -\pi/4)]^{3/2}} 
\Theta \left(|\sin (\phi - \pi/4)| - \frac{|\Omega|}{v_F K} \right)\,.
\end{split}
\end{equation}

The functions $f_1$,  $f_2$ and $f_3$ are obtained from
(\ref{kinetic.frequency.Landau}), (\ref{kinetic.momentum.Landau}) and 
(\ref{kinetic.frequency.momentum.Landau}),
respectively. Deriving (\ref{f1}) - (\ref{f3}) we used the 
the assumption $|\Omega|/P_j \ll 1$ replacing the square roots in 
(\ref{kinetic.frequency.Landau}), (\ref{kinetic.momentum.Landau}) and
(\ref{kinetic.frequency.momentum.Landau}) by 1. Furthermore,
we kept only $\sin^2 \psi_j$ from (\ref{kinetic.frequency.Landau}). 
Nevertheless, we  kept the $\Theta$-functions which are present 
in (\ref{kinetic.frequency.Landau}), (\ref{kinetic.momentum.Landau}) and
(\ref{kinetic.frequency.momentum.Landau}) because they are 
essential in imposing the condition $\Omega/P_j$. 

\begin{figure}[h]
\centering{
\includegraphics[width=8cm]{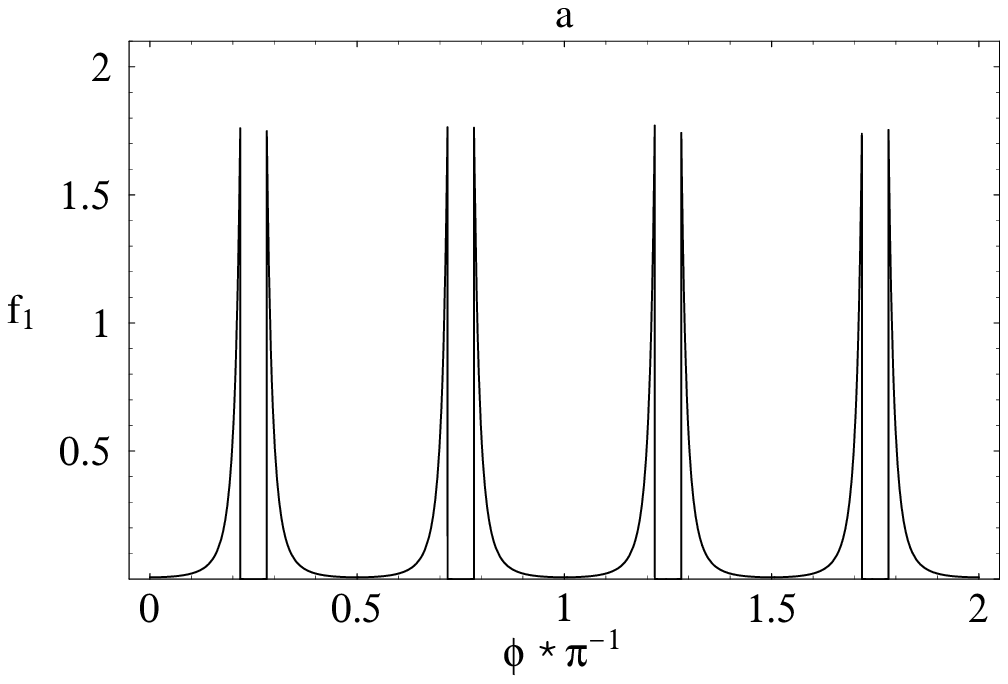}
\includegraphics[width=8cm]{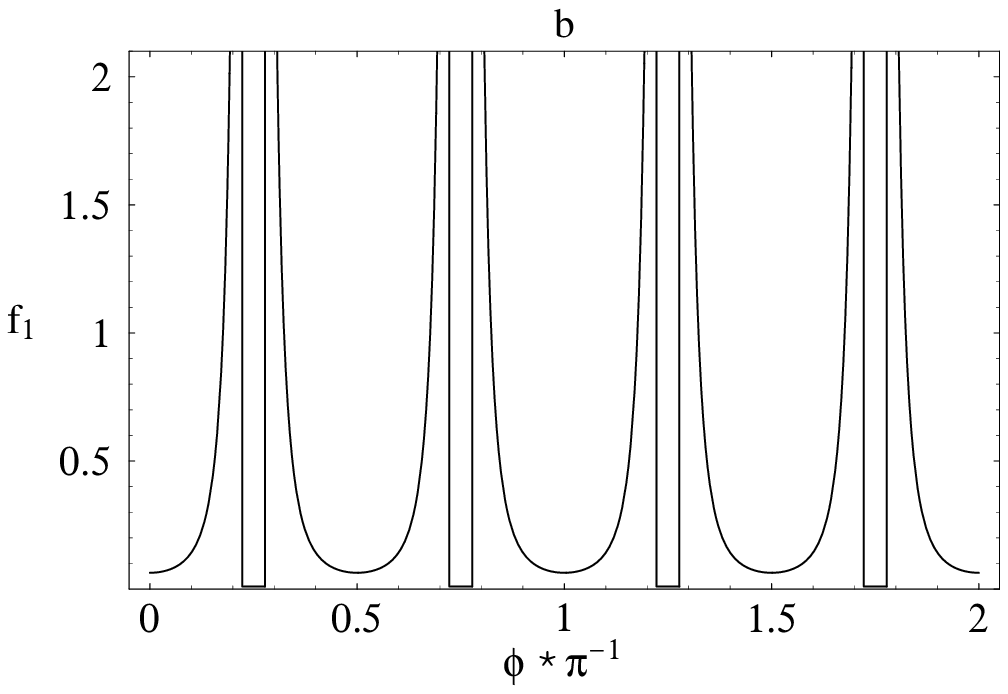}}
\caption{(a) The angular dependence, 
$f_1 \left(\frac{\Omega}{v_F K}, \phi \right)$ given by Eq.~(\ref{f1})
of the Landau term $\sim \Omega^3/K$ for $\alpha_D =20$
and $|\Omega|/v_F K =0.1$. (b) The same function calculated directly 
from Eq.~(\ref{kinetic.frequency.Landau}).}  
\label{fig:4}
\end{figure}

\begin{figure}[h]
\centering{
\includegraphics[width=8cm]{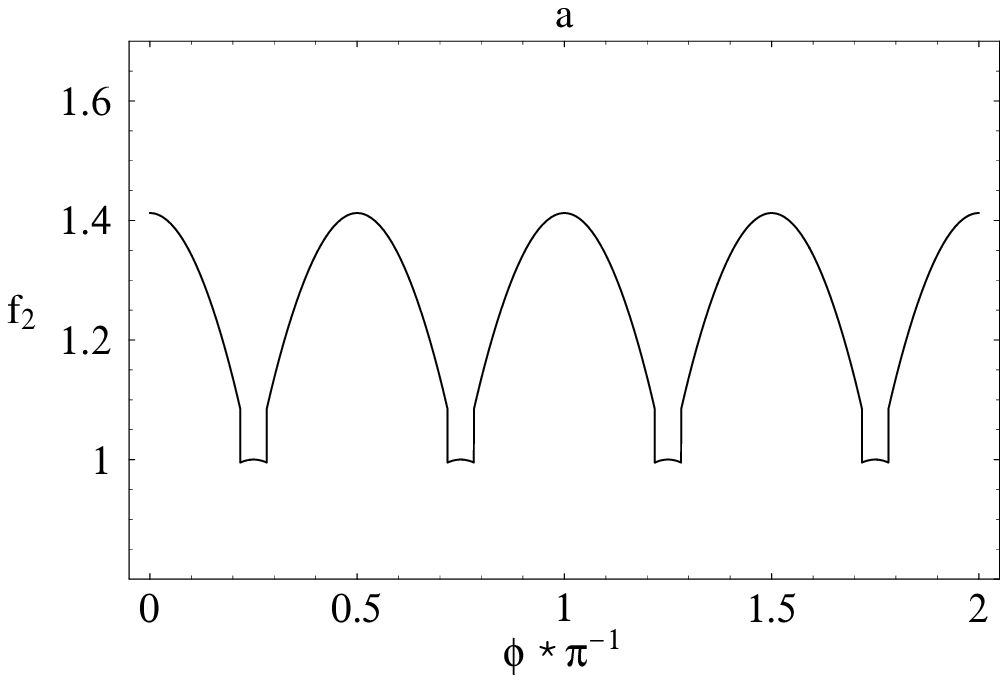}
\includegraphics[width=8cm]{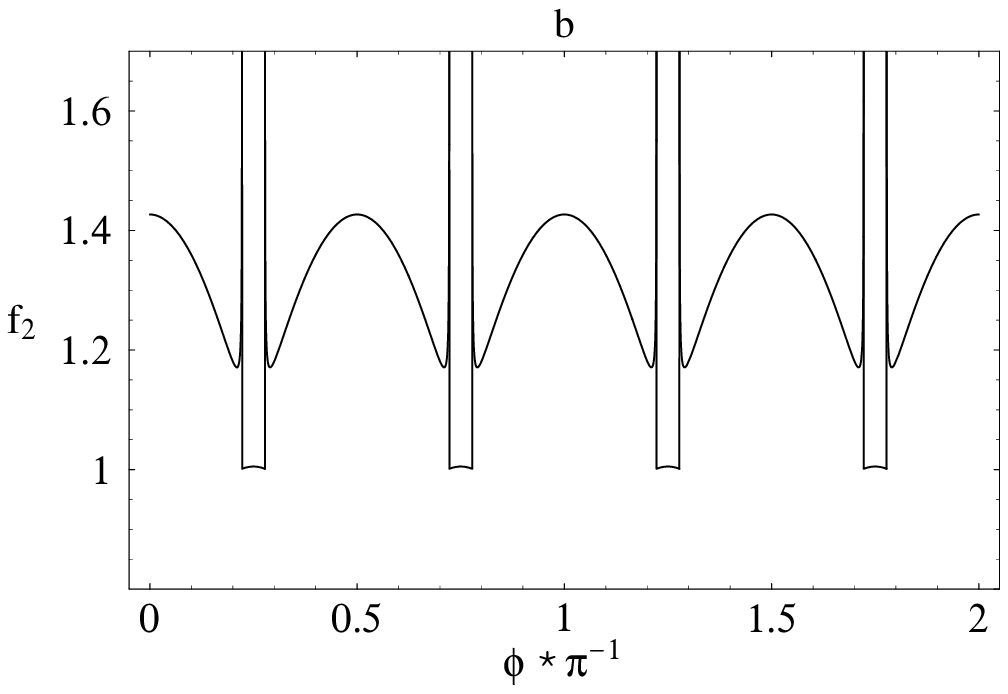}}
\caption{(a) The angular dependence, 
$f_2 \left(\frac{\Omega}{v_F K}, \phi \right)$ given by Eq.~(\ref{f2})
of the Landau term $\sim \Omega K$. The parameters $\alpha_D$
and $|\Omega|/v_F K$ are the same as in Fig.~\ref{fig:4}.
(b)  The same function calculated directly 
from Eq.~(\ref{kinetic.momentum.Landau}).}  
\label{fig:5}
\end{figure}

\begin{figure}[h]
\centering{
\includegraphics[width=8cm]{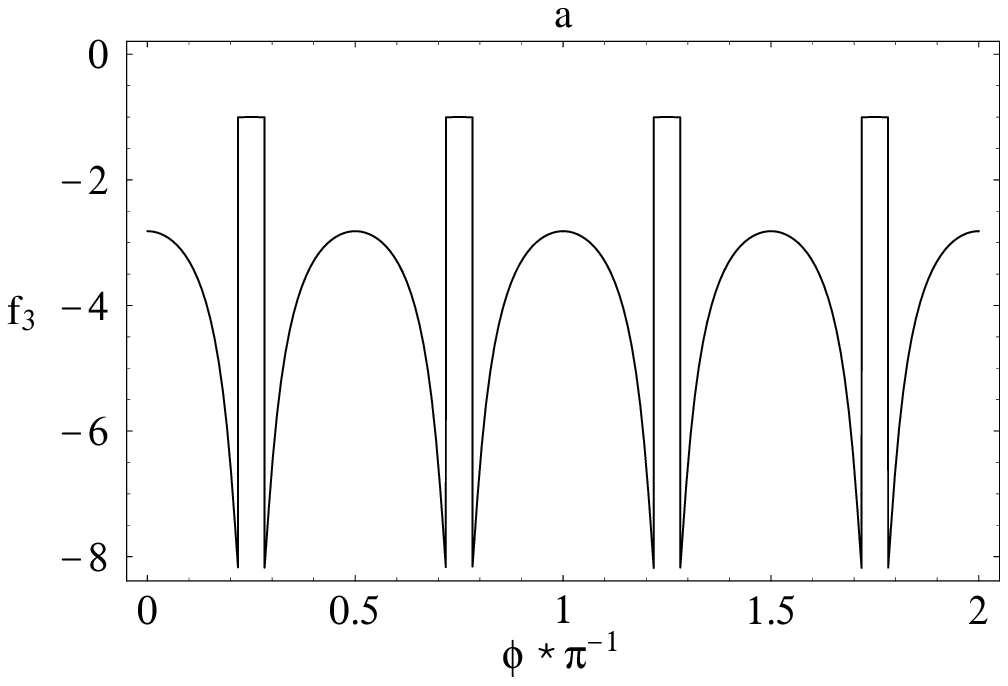}
\includegraphics[width=8cm]{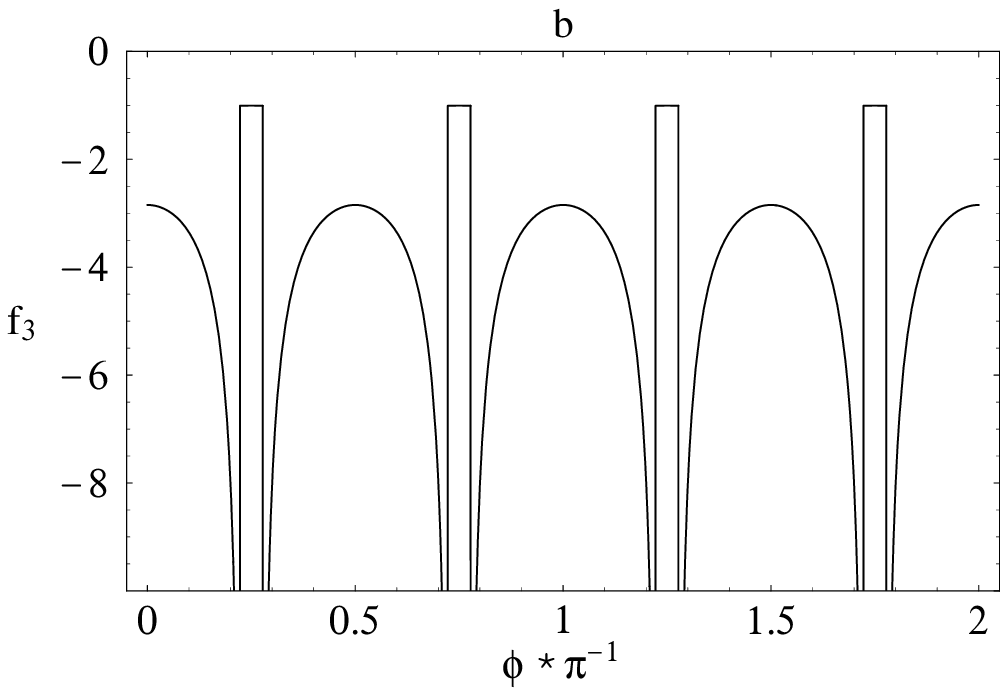}}
\caption{(a) The angular dependence, 
$f_3 \left(\frac{\Omega}{v_F K}, \phi \right)$ given by Eq.~(\ref{f3})
of the Landau term $\sim \Omega^3/ K$. The parameters $\alpha_D$
and $|\Omega|/v_F K$ are the same as in Fig.~\ref{fig:4}.
(b)  The same function calculated directly 
from Eq.~(\ref{kinetic.frequency.momentum.Landau}).}  
\label{fig:6}
\end{figure}

In Figs.~\ref{fig:4}~(a), \ref{fig:5}~(a) and \ref{fig:6}~(a) 
we show the functions $f_1$, $f_2$ and $f_3$ which describe the intensity
of Landau damping ($f_1$ and $f_3$ for the term $\sim \Omega^3/K$
and $f_2$ for the term $\sim \Omega K$, respectively) as a function
of the direction in the plane. Although the functions $f_1$ and $f_3$
describe the directional dependence for the same $\sim \Omega^3/K$
term, we do not combine these functions into the single function
because they originate from the different expressions.
(We recall that $f_1$ originates from Eq.~(\ref{kinetic.frequency.Landau})
and $f_3$ originates from Eq.~(\ref{kinetic.frequency.momentum.Landau}),
respectively.) 
For the comparison in Figs.~\ref{fig:4}~(b), \ref{fig:5}~(b) 
and \ref{fig:6}~(b) we show the directional dependences
calculated by the direct substitution of Eq.~(\ref{transform}) into
Eqs.~(\ref{kinetic.frequency.Landau}), (\ref{kinetic.momentum.Landau}) 
and (\ref{kinetic.frequency.momentum.Landau}) without making the 
approximations we just described. As one can see, the approximate
representation (\ref{effective.action.singular}), (\ref{f1}) - (\ref{f3})
despite its relatively simple form (we have used only the terms
$\sim \Omega^3/K$ and $\sim \Omega K$) gives reasonably good
expression for the Landau damping terms. The biggest discrepancies
are seen between Fig.~\ref{fig:4}~(a) and Fig.~\ref{fig:4}~(b)
because more approximations were made to obtain $f_{1}$ term.

As we mentioned in Sec.~\ref{sec:Landau.damping}, the angular dependence
described by $f_1$ (see Fig.~\ref{fig:4}~(a))
coincides with the angular dependence of the
ultrasonic attenuation \cite{Carbotte} in the limit $\Omega/v_{F} K \to 0$.
Indeed the $\Theta$-functions from (\ref{f1}) disappear when
$\Omega/v_{F} K \to 0$ and  Eq.~(\ref{f1}) reduces to the
corresponding equation from \cite{Carbotte}. 

We stress also
that the analytical structure of the damping terms in
(\ref{effective.action.singular}) appears to be quite different
from the dissipation introduced in
\cite{Benfatto.dissipation} into the ``phase only'' action basing
on the low frequency conductivity measurements.

Since $f_1+2f_3$ and $f_2$ terms are odd functions of the frequency $\Omega$,
they integrate to zero in $\Omega_{kin}$.
Nevertheless, the damping terms are manifest in the equation
of motion and in the propagator of the BKT mode which is
considered below.

Comparing Figs.~\ref{fig:4}, \ref{fig:5}  and \ref{fig:6}, one can see that    
the functions $f_1$ and $f_3$ have more pronounced angular dependence 
than $f_2$. Although as we explained above, the contribution from a 
given node is zero if for some ${\bf K}$ the condition
(\ref{condition.approximate}) is not satisfied, this 
condition can still simultaneously be satisfied for the neighboring
nodes, so that the sum over all nodes in $f_1$ -- $f_3$ is not necessarily
zero.
We note also that by the absolute value both $f_1$ (or $f_3$) and $f_2$
terms are practically the same. This can be easily seen if we rewrite,
for example, $f_1$ term as
\begin{equation}\label{f1.rewrite}
\frac{\Omega^3}{K} \frac{\ln 2}{\pi}
\frac{T}{v_{\Delta }v_{F}^{2}} f_1 \left(\frac{\Omega}{v_F K}, \phi \right) =
\Omega K \left(\frac{\Omega}{v_F K}\right)^2 \frac{\ln 2}{\pi} 
\frac{T}{v_{\Delta}}
f_1  \left(\frac{\Omega}{v_F K}, \phi \right)
\approx \Omega K  \frac{\ln 2}{6} \frac{T}{v_F} 
f_1 \left(\sqrt{\frac{\pi}{6 \alpha_{D}}}, \phi \right) \,,
\end{equation}
where writing the last identity we explicitly used the ratio
$\Omega/v_F K$ for the BKT mode at $T =0$ given above.

\begin{figure}[h]
\centering{
\includegraphics[width=10cm]{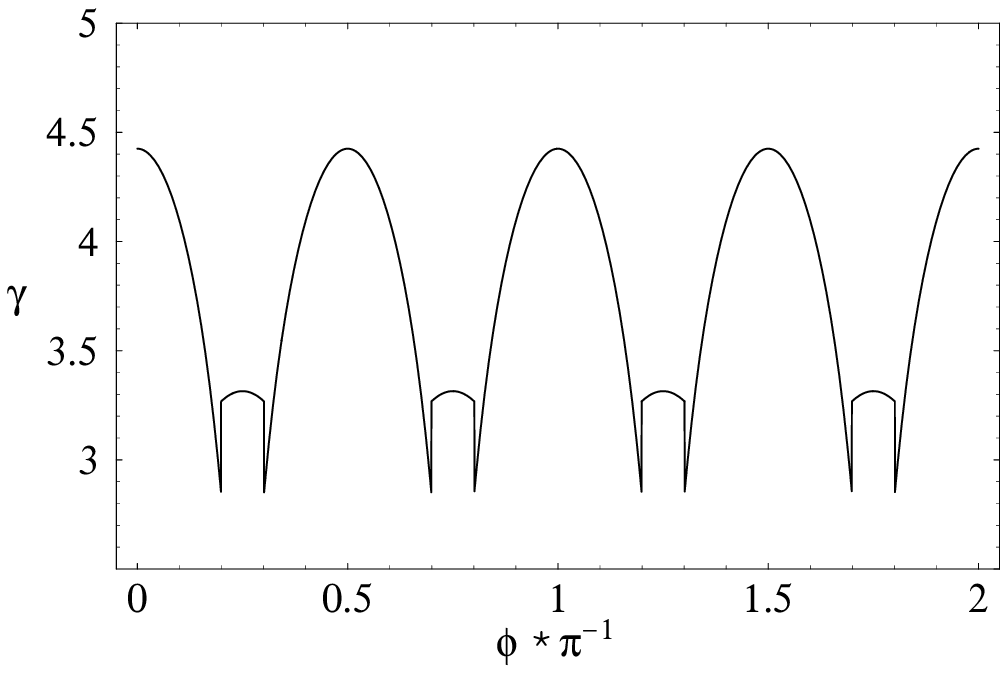}}
\caption{The angular dependence of the decay constant 
$\gamma \left(\frac{\Omega}{v_F K}, \phi \right)$
in $\theta$-propagator (\ref{propagator}) for 
$\alpha_D = 20$ and $|\Omega|/v_F K = \sqrt{\frac{\pi}{6 \alpha_{D}}} 
\simeq 0.16$.}  
\label{fig:7}
\end{figure}

In such a way following \cite{Aitchison} we can  write an approximate 
propagator of the BKT mode near $T =0$ as
\begin{equation}
\label{propagator}
D_{\theta}^{R}(\Omega, \mathbf{K}) \approx 
\left\{ \frac{1}{4 a \sqrt{\pi v_{F} v_{\Delta} }} 
\left[ \Omega^2 - K^2 
\left( \frac{\pi v_{F} v_{\Delta}}{6} - \frac{2 \ln2 a v_{F}}{\pi}
\sqrt{ \alpha_{D}} T  \right) + 2 i a \gamma (\phi) T \Omega  K
\right] \right\}^{-1}
\end{equation}
with
\begin{equation}
\label{gamma}
\gamma ( \phi ) =
\ln 2 \left[\frac{1}{6} \sqrt{\frac{\pi}{\alpha_{D}}}
\left(f_1  \left(\sqrt{\frac{\pi}{6 \alpha_{D}}}, \phi \right) 
+ 2 f_3  \left(\sqrt{\frac{\pi}{6 \alpha_{D}}}, \phi \right) \right)
+ \sqrt{\frac{\alpha_{D}}{\pi}} 
f_2 \left(\sqrt{\frac{\pi}{6 \alpha_{D}}}, \phi \right)
\right]\,.
\end{equation}
As discussed in \cite{Aitchison} Eq.~(\ref{propagator}) has
the form of a bosonic propagator with damping and its line-width
has an explicit $\mathbf{K}$ dependence. In contrast to
the $s$-wave case,  the width depends not only on the absolute value of 
$|\mathbf{K}|$, but also on the direction of $\mathbf{K}$. 
The angular dependence for $\gamma(\phi)$
is shown in Fig.~\ref{fig:7}. 
%The dependence appears to be rather smooth, so that in practice 
%$\gamma$ can be regarded as a constant which does not depend on $\phi$.
Despite a simple form of the propagator (\ref{propagator}),
the corresponding effective wave operator (the transform of  
$D_{\theta}^{-1}$ to space and time)
has a nonlocal form in coordinate space due to the presence
of the damping term which depends \cite{Aitchison} on $|\mathbf{K}|$. 
%To avoid this problem one can remain in ``Matsubara'' frequency space.

Comparing the second term in parentheses of (\ref{propagator}) with
the damping term one can see that they have the same order of magnitude 
showing that even for the low temperature region the Landau damping becomes 
important and its magnitude is comparable with term which describes the 
linear low temperature decrease in the superfluid stiffness.

We stress also the difference between the Landau damping representation
in Eqs.~(\ref{propagator}) and (\ref{effective.action.singular}).
While the propagator (\ref{propagator}) relies on the particular 
dispersion law for the BKT mode with the approximate $v$
given by Eq.~(\ref{BKTmode.d}), our Eq.~(\ref{effective.action.singular})
along with Eqs.~(\ref{f1}) - (\ref{f3}) present a rather general
representation for the Landau damping terms derivation of which
did not rely on any particular dispersion law for the phase
excitations. Thus it can be, in principle, used to describe the damping
of the plasmons in charged superconductor. The only assumption we made is that 
$\Omega/v_{F} K \ll 1$, which was used to derive more simple and transparent
representation for the Landau damping. For the more general case
one should use the results from Sec.~\ref{sec:Landau.damping}.

\section{Concluding Remarks} 
\label{sec:conclusion}

It is very important to note that the collective phase excitations
described by the propagator similar to (\ref{propagator}) 
can and have indeed been studied experimentally. Indeed 
the measurements of the order parameter
dynamical structure factor in the dirty Al films allowed
to extract the dispersion relation of the corresponding Carlson-Goldman
mode and to investigate its temperature dependence \cite{Goldman}. 
It is important to stress that since the real systems are 
{\em charged} this mode appears to be different from
the sound-like Bogolyubov-Anderson mode which exists in a 
{\em neutral} superfluid Fermi liquid. However, as discussed in 
\cite{Kulik} (see also \cite{Takada} and Refs. therein)
the discovery of Carlson-Goldman collective mode \cite{Goldman}
overcame the widespread opinion that the Bogolyubov-Anderson
collective oscillations predicted for a neutral superfluid should 
be forced to the frequencies on the order of the plasma frequency.
It appears that if under certain conditions the frequency of the phase 
excitations is sufficiently low, the normal electron fluid may screen 
completely the associated electric field, thereby making the theory of 
uncharged superconductor applicable \cite{Brieskorn}. The conditions
for screening and occurrence of the Carlson-Goldman mode in $s$-wave
superconductors are favorable only if the systems are dirty
(to suppress the Landau damping) and $T \lesssim T_{c}$ \cite{Kulik}.
As recently claimed in \cite{Takada} in $d$-wave superconductors
the conditions appear to much less strict, so that the Carlson-Goldman
mode may be observed in the clean systems for $T$ down to $0.2 T_{c}$.
Now there are some new questions which would be interesting to address 
from both experimental and theoretical points of view.

\noindent
First of all, there is a general question whether the Carlson-Goldman
mode which is the equivalent of the Goldstone mode for the charged
system can be observed experimentally. There are also a lot
of theoretical questions which study would make this kind of experiment
possible. Leaving aside the specific of the Carlson-Goldman mode
studied in \cite{Takada} we mention some of them which are directly
related to the present work.     

\noindent
1. The Carlson-Goldman experiment \cite{Goldman} measures 
so called dynamical structure factor. This factor has a peak 
associated with the phase excitations the width of which is primarily 
controlled by the Landau damping. Therefore our result that 
the intensity of Landau damping has a strong directional 
dependence should be taken into account in the calculation of
the structure factor. In particular, the lowest for the observation
of the Carlson-Goldman mode temperature $0.2 T_c$ in \cite{Takada}
is obtained for the nodal direction. As can be seen from, for example,
Fig.\ref{fig:4} and \ref{fig:5}, the corresponding 
Landau damping terms become maximal in the vicinity
of this direction which should likely result in widening
of the corresponding peak. If this widening is too big that the peak
may become unobservable, this would again suggest that
the dirty samples should be used for the experiment.
    
\noindent
2. Taking into account the lattice effects and the dominance
of the nodal excitations we have obtained that the velocity
of the phase excitations at $T \to 0$ is given by
$v_{\rm lat} = v_{F} \sqrt{\pi /6 \alpha_{D}}$
(see Eq.~(\ref{BKTmode.d}) and the 
discussion after Eq.~(\ref{kinetic.1.final})).
This expression appears to be quite different from the well-known continuum
result, $v_{\rm cont} = v_F/\sqrt{2}$. 
The estimates of the velocities  $v_{\rm cont}$ and
$v_{\rm lat}$ obtained from the values $v_F$ and $\alpha_{D}$ 
\cite{Chiao} are presented in Table~\ref{tab1}. 
It is seen from these estimates that the difference between
the description proceeding from continuum and lattice models can be
quite different. However the estimates of the minimal 
for the observation of the Carlson-Goldman mode temperature in \cite{Takada}
are based on the continuum expression $v_{\rm cont} = v_F/\sqrt{2}$. 
Thus it would be interesting to reconsider these estimates
for the lattice case.

\noindent
3. Finally, as pointed out in \cite{Randeria.action}
if the plasmon is at finite frequency at $\mathbf{K} \to 0$, the Landau
damping does not occur since when $\Omega$ is finite and $\mathbf{K} \to 0$
it is impossible to satisfy  {\v C}erenkov condition discussed
above. However, a very small value of the plasma frequency in HTSC
suggests that the Landau damping may still be relevant when $\mathbf{K}$
is nonzero and the {\v C}erenkov condition can be satisfied.
Thus the corresponding damping terms should be included in the
``phase only'' actions which are used to describe plasma excitations
in  $d$-wave superconductors. Predicted anisotropy of the Landau
damping would result in the damping anisotropy for the plasma
excitations.  

To summarize, we have considered the ``phase only'' effective action
for so called {\it neutral (or uncharged)\/} fermionic
superfluid with $d$- and $s$-wave pairing in 2D. When the damping terms
are included into this action, it turns out non-local in coordinate
space and its analytical structure for the $d$-wave case is very
different from the $s$-wave case. 
To consider a charged superfluid it is necessary to combine the approaches 
used in the present paper and in \cite{Randeria.action,Takada} to 
consider the Landau damping in the presence of Coulomb interaction.

\section{Acknowledgments}

We gratefully acknowledge Dr. E.V.~Gorbar, Dr. M.~Capezzali,
Prof. V.E.~Mkrtchyan and especially 
Prof. V.P.~Gusynin for careful reading of the manuscript and valuable 
suggestions. We  thank Prof.~P.~Martinoli and Dr.~X.~Zotos for stimulating
discussion and for bringing the papers \cite{Takada,Goldman,Brieskorn} 
to our attention.
S.G.Sh. is  grateful to the members of the Institut de Physique, 
Universit\'e de Neuch\^atel for hospitality.
This work was supported by the research project 2000-061901.00/1 and
by the SCOPES-project 7UKPJ062150.00/1
of the Swiss National Science Foundation.
V.M.L. also thanks NATO grant CP/UN/19/C/2000/PO (Program OUTREACH) 
for partial support and Centro de Fisica do Porto (Portugal) for hospitality.

\appendix
\section{}
\label{sec:A}

Substituting (\ref{Green.neutral.momentum}) into (\ref{kinetic.frequency}) and
(\ref{kinetic.momentum.general}), using the definitions for $\pi_{ij}$
and $\Pi_{ij}$ and evaluating the matrix traces for the
diagonal terms we obtain
\begin{equation}
\label{tr}
\begin{split}
& A_{\pm} \equiv
\begin{bmatrix}
\Pi_{00}^{\alpha \beta}(i \Omega_{n}, {\bf K}) \\
\Pi_{33}(i \Omega_{n}, {\bf K})
\end{bmatrix} = \\
\\ &  T \sum_{l = - \infty}^{\infty} \int \frac{d^2 k}{(2 \pi)^2}\,
\frac{2[(i \omega_l + i \Omega_n) i \omega_l + \xi({\bf k} - {\bf
K}/2)\xi({\bf k} + {\bf K}/2) \pm \Delta({\bf k} - {\bf K}/2)\Delta({\bf k} + {\bf
K}/2)] V_{\pm}(\mathbf{k}) }
{[(i \omega_l + i \Omega_n)^2-\xi^2({\bf k} + {\bf K}/2)-\Delta^2({\bf k}
+ {\bf K}/2)][(i \omega_l)^2-\xi^2({\bf k} - {\bf K}/2)-\Delta^2({\bf k} - {\bf
K}/2)]}\,, \\ 
& \qquad \qquad \qquad V_{\pm}(\mathbf{k}) \equiv 
\begin{bmatrix}
v_{F \alpha}(\mathbf{k}) v_{F \beta}(\mathbf{k}), & ``+''; \\
1, & ``-''.
\end{bmatrix}
\end{split}
\end{equation}
Performing in (\ref{tr}) the summation over Matsubara frequencies and
simplifying the result we arrive at
\begin{equation}\label{A}
\begin{split}
& A_{\pm} =  \\
&  - \int \frac{d^2 k}{(2 \pi)^2}  \left\{ \frac{1}{2} \left(1 -
\frac{\xi_{-}\xi_{+} \pm \Delta_{-} \Delta_{+} }{E_{-}E_{+}} \right) \left[
\frac{1}{E_{+} + E_{-} + i \Omega_n} +
\frac{1}{E_{+} + E_{-} - i \Omega_n}\right] [1 - n_F(E_{-}) - n_F(E_{+})]
\right. \\
& \left. + \frac{1}{2} \left(1 + \frac{\xi_{-}\xi_{+} \pm \Delta_{-} \Delta_{+}
}{E_{-}E_{+}} \right)\left[ \frac{1}{E_{+} - E_{-} + i \Omega_n} +
\frac{1}{E_{+} - E_{-} - i \Omega_n}\right] [n_F(E_{-}) - n_F(E_{+})]
\right\}V_{\pm}(\mathbf{k})\,,
\end{split}
\end{equation}
where $\xi_{\pm} \equiv \xi({\bf k} \pm {\bf K}/2)$, $E_{\pm} \equiv E({\bf k}
\pm {\bf K}/2)$ and $\Delta_{\pm} \equiv \Delta({\bf k} \pm {\bf K}/2)$. The second
term in Eq.~(\ref{A}) can be further simplified if one notices that the  term
with $1/(E_{+} - E_{-} - i \Omega_n)$ transforms into the term $(-1)/(E_{+} -
E_{-} + i \Omega_n)$ when $\mathbf{k} \to -\mathbf{k}$. Note that the
$A_{\pm}$ terms do coincide with the corresponding terms in \cite{Aitchison}.

For the mixed term (\ref{kinetic.frequency.momentum.general}) one obtains
\begin{equation}\label{tr.mixed}
\begin{split}
& \Pi_{03}^{\alpha}(i \Omega_{n}, {\bf K}) =\\ 
& T \sum_{l = - \infty}^{\infty} \int \frac{d^2 k}{(2 \pi)^2}\,
\frac{2[(i \omega_l + i \Omega_n)  \xi({\bf k} - {\bf K}/2) + \xi({\bf k} +
{\bf K}/2) i \omega_l] v_{F \alpha}(\mathbf{k})}
{[(i \omega_l + i \Omega_n)^2-\xi^2({\bf k} + {\bf
K}/2)-\Delta^2({\bf k} + {\bf K}/2)][(i \omega_l)^2-\xi^2({\bf k} - {\bf
K}/2)-\Delta^2({\bf k} - {\bf K}/2)]} \\
& =  \int \frac{d^2 k}{(2 \pi)^2} \left\{ \left(\frac{\xi_{+}}{2E_{+}} -
\frac{\xi_{-}}{2E_{-}} \right) \left[ \frac{1}{E_{+} + E_{-} + i \Omega_n} -
\frac{1}{E_{+} + E_{-} - i \Omega_n}\right] [1 - n_F(E_{-}) - n_F(E_{+})]
\right. \\
& \left. + \left(\frac{\xi_{+}}{2E_{+}} + \frac{\xi_{-}}{2E_{-}} \right) \left[
\frac{1}{E_{+} - E_{-} + i \Omega_n} - \frac{1}{E_{+} - E_{-} - i
\Omega_n}\right] [n_F(E_{-}) - n_F(E_{+})] \right\} 
v_{F \alpha}(\mathbf{k}) \\
& =  \int \frac{d^2 k}{(2 \pi)^2} \left\{ \frac{\xi_{+}}{E_{+}}  
 \left[ \frac{1}{E_{+} + E_{-} + i \Omega_n} -
\frac{1}{E_{+} + E_{-} - i \Omega_n}\right] [1 - n_F(E_{-}) - n_F(E_{+})]
\right. \\
& \left. + \left(\frac{\xi_{+}}{E_{+}} + \frac{\xi_{-}}{E_{-}} \right) 
\frac{1}{E_{+} - E_{-} + i \Omega_n} 
[n_F(E_{-}) - n_F(E_{+})] \right\} v_{F \alpha}(\mathbf{k}) \,.
\end{split}
\end{equation}
To get the last identity we again replaced $\mathbf{k} \to -\mathbf{k}$
and took into account that $v_{F}(\mathbf{k})$ also changes sign
under this transformation. 
One can also check that  $\Pi_{30}^{\alpha}(i\Omega_{n},{\bf K}) = 
\Pi_{03}^{\alpha} (i\Omega_{n},{\bf K})$.
Eq.~(\ref{tr.mixed}) is also in agreement with  the corresponding term
in \cite{Aitchison} denoted as $D$. 

The first and second terms in (\ref{A}) and (\ref{tr.mixed}) have a  
clear physical interpretation \cite{Schrieffer}. 
The first term gives the contribution
from ``superfluid'' electrons. The second term gives the contribution
of the thermally excited quasiparticles (i.e. ``normal'' fluid component).
The essential physical difference between the two terms is that
the superfluid term involves creation of two quasiparticles,
with the minimum excitation energy in the $s$-wave case being $2 \Delta_s$.
(For the $d$-wave case the minimum energy turns out to be zero in the
four nodal points, see the discussion at the end of 
Sec.~\ref{sec:Landau.damping}.) 

On the other hand, the normal fluid term
involves scattering of the quasiparticles {\it already present}
and the excitation energy in this case can be arbitrary small,
as in the normal metal, independently whether we are in the vicinity of 
the node or not.  Of course, the number of the 
quasiparticles participating in this scattering depends on
the value of the gap and drastically increases if the gap is equal
to zero in some points.
As we shall see, the Landau damping terms originate from the second,
normal fluid term. Note that our Eqs.~(\ref{A}) and (\ref{tr.mixed})
(as well as Eqs.~ (\ref{kinetic.frequency}),
(\ref{kinetic.momentum.general}) and
(\ref{kinetic.frequency.momentum.general}))
are suitable for studying both $s$- and $d$-wave cases.

%\iffalse

%\newpage

\begin{table}[h]
\caption{The Fermi velocities $v_F$ and anisotropies of the Dirac
spectrum, $\alpha_{D}$ of YBa$_2$Cu$_3$O$_{6.9}$ and
Bi$_2$Sr$_2$CaCu$_2$O$_8$ at optimal doping from \cite{Chiao}. 
The velocities of the phase excitations for the continuum, $v_{\rm cont}$ and 
lattice, $v_{\rm lat}$ limits at $T=0$.}
\vspace{5mm}
\begin{tabular}{|r|c|c|c|c|}
\hline \hline 
& $v_{F}$, $\times 10^7$ cm/s & $\alpha_{D}$ & 
$v_{\rm cont}$, $\times 10^7$ sm/s & $v_{\rm lat}$, $\times 10^6$ sm/c\\
\hline YBCO  & 2.5 & 14 & 1.77  & 4.8 \\
\hline BSCCO & 2.5 & 19 & 1.77  & 4.2 \\
\hline \hline
\end{tabular}
\label{tab1}
\end{table}

%\fi


\begin{thebibliography}{99}
\bibitem{foot1} We stress that this 
field is Grassmann variable since it satisfies the appropriate 
anti-commutation algebra (see Refs. in \cite{Loktev.review}).
The transformation (\ref{fermi.parametrization}) can be also regarded
as a gauge transformation.
\end{thebibliography}
\end{document}